\documentstyle[epsfig,11pt]{article}
\newcommand{\zp}[3]{Z. Phys.\ C#1 (19#2) #3}
\newcommand{\pl}[3]{Phys.\ Lett.\ #1B (19#2) #3}
\newcommand{\np}[3]{Nucl.\ Phys.\ B#1 (19#2) #3}
\newcommand{\prd}[3]{Phys.\ Rev.\ D#1 (19#2) #3}
\newcommand{\prl}[3]{Phys.\ Rev.\ Lett.\ #1 (19#2) #3}

\newcommand{\md}{\mbox{d}}
\newcommand{\beq}{\begin{equation}}
\newcommand{\eeq}{\end{equation}}
\newcommand{\bea}{\begin{eqnarray}}
\newcommand{\eea}{\end{eqnarray}}

\newcommand{\Li}{\mbox{Li}}
\newcommand{\nlf}{\mbox{$n_{\mbox{\scriptsize{}lf}}$}}
\newcommand{\nf}{\mbox{$n_{\mbox{\scriptsize{}f}}$}}
\newcommand{\mur}{\mbox{$\mu_{R}$}}
\newcommand{\mursq}{\mbox{$\mu_{R}^{2}$}}
\newcommand{\alps}{\mbox{$\alpha_{\mbox{\scriptsize s}}$}}
\newcommand{\alpsq}{\mbox{$\alpha_{\mbox{\scriptsize s}}^{2}$}}
\newcommand{\alptr}{\mbox{$\alpha_{\mbox{\scriptsize s}}^{3}$}}
\def\simgt{\rlap{\lower 3.5 pt \hbox{$\mathchar \sim$}}
                                \raise 1pt \hbox {$>$}}
\def\simlt{\rlap{\lower 3.5 pt \hbox{$\mathchar \sim$}}
                                \raise 1pt \hbox {$<$}}
\catcode`\@=11
\def\@citex[#1]#2{\if@filesw\immediate\write\@auxout{\string\citation{#2}}\fi
  \def\@citea{}\@cite{\@for\@citeb:=#2\do
    {\@citea\def\@citea{,\penalty\@m}\@ifundefined
       {b@\@citeb}{{\bf ?}\@warning
       {Citation `\@citeb' on page \thepage \space undefined}}%
\hbox{\csname b@\@citeb\endcsname}}}{#1}}

\def\citer{\@ifnextchar [{\@tempswatrue\@citexr}{\@tempswafalse\@citexr[]}}

\def\@citexr[#1]#2{\if@filesw\immediate\write\@auxout{\string\citation{#2}}\fi
  \def\@citea{}\@cite{\@for\@citeb:=#2\do
    {\@citea\def\@citea{--\penalty\@m}\@ifundefined
       {b@\@citeb}{{\bf ?}\@warning
       {Citation `\@citeb' on page \thepage \space undefined}}%
\hbox{\csname b@\@citeb\endcsname}}}{#1}}
\relax
\begin{document}

\thispagestyle{empty}

\hfill\vbox{\hbox{\bf DESY 95-155}
                                   }
\vspace{1.0in}

\begin{center}
\boldmath
{\Large\bf QCD Corrections to Inelastic $J/\psi$ Photoproduction} \\
\unboldmath

\vspace{1.5cm}

{\large \sc Michael~Kr\"amer}$\,^\star$

\vspace{0.3cm}

Deutsches Elektronen-Synchrotron DESY\\[1mm]
D-22603 Hamburg, Germany.

\end{center}

\vspace{1.5cm}

\noindent
\begin{abstract}
We present a complete calculation of the higher-order perturbative
QCD corrections to inelastic photoproduction of $J/\psi$ particles.
A comprehensive analysis of total cross sections and differential
distributions for the energy range of the fixed-target experiments
and for inelastic $J/\psi$ photoproduction at HERA is performed.
The cross section and the $J/\psi$ energy spectrum are compared
with the available photoproduction data including first results
from HERA. This analysis will not only provide information on the
gluon distribution of the proton but appears to be a clean test for
the underlying picture of quarkonium production as developed so far
in the perturbative QCD sector.
\end{abstract}

\vfill

\noindent
{\small $\star\,$ E-mail: mkraemer@desy.de}
\newpage

\section{Introduction}
The measurement of the gluon distribution in the nucleon is one
of the important goals of lepton-nucleon scattering experiments.
The classical methods exploit the evolution of the nucleon structure
functions with the momentum transfer and the size of the longitudinal
structure function.  With rising energies, however, jet physics and
the production of heavy quark states become important complementary
tools. Heavy flavour production in lepton-nucleon scattering is
dominated by photon-gluon fusion and can thus yield direct information
on the gluon distribution in the nucleon $G(x,Q^2)$. Besides open charm
and bottom production, the formation of $J/\psi$ bound states in
inelastic photoproduction experiments
\begin{equation}
\gamma + {\cal N} \to J/\psi + X
\end{equation}
provides an experimentally attractive method since $J/\psi$ particles
are easy to tag in the leptonic decay modes.

The production of heavy quarks in high energy photon-proton collisions
can be calculated in perturbative QCD. The mass of the heavy quark,
$m_Q \gg \Lambda_{\mbox{\scriptsize QCD}}$, acts as a cutoff and sets
the scale for the perturbative calculations \cite{CSS86}.  However,
the subsequent transition from the colour-octet $Q\overline{Q}$ pair
to a physical quarkonium state introduces non-perturbative aspects.
Two different mechanisms of bound state formation have been employed
in previous analyses (for a recent review see Ref.\cite{GS94}):

\noindent
 (i) The {\em local duality approach} \cite{LOCAL_DUALITY} assumes
 that the colour-octet $Q\overline{Q}$ pair rearranges itself into
 a colour-singlet bound state by the emission of non-perturbative
 soft gluons. According to the arguments of semi-local duality,
 one averages over all possible quarkonia states by integrating
 the perturbative cross section for inclusive $Q\overline{Q}$
 production from the quark threshold ($=2m_Q$) to the physical
 threshold for the production of a pair of heavy-light mesons
 ($=2m_D$ for the $c\bar{c}$ system). The probability to generate
 a particular state, e.g. $J/\psi$, depends on different dynamical
 details of the production mechanism and cannot be absolutely predicted
 in this model. Another serious drawback of the duality approach is the
 fact that higher-order QCD corrections cannot be included since there
 is no unique way to decide what part of the radiatively emitted gluons
 are to be considered as a part of the bound state. Although dual models
 might describe some qualitative features of quarkonium production, they
 do not allow for a quantitative prediction and will therefore not be
 discussed in the present context.

\noindent
 (ii) In the {\em colour-singlet(CS) mechanism} \citer{BJ81,JGK82}
 the quarkonium state is described by a colour-singlet $Q\overline{Q}$
 pair with the appropriate spin, angular-momentum and charge conjugation
 quantum numbers. In the static approximation, in which the
 motion of the charm quarks in the bound state is neglected, the
 production cross section factorizes into a short distance matrix
 element which describes the production of a $Q\overline{Q}$ pair
 within a region of size $1/m_Q$, and a long distance factor that
 contains all the nonperturbative dynamics of the bound state formation.
 The short distance cross section can be calculated as a perturbative
 expansion in powers of the strong coupling constant $\alps(m_Q)$,
 evaluated at a scale set approximately by the heavy quark mass, while
 the long-distance factor is related to the leptonic width.

A rigorous framework for treating quarkonium production and decays
has recently been developed in Ref.\cite{BBL95} (see also
Ref.\cite{MS95}). The so-called factorization scheme is based on
the use of non-relativistic QCD (NRQCD) \cite{CL86} to separate the
short distance parts from the long-distance matrix elements. The
factorization approach explicitly takes into account the complete
structure of the quarkonium Fock space.  For the production of
$S$-wave quarkonia, like $J/\psi$, the colour-octet matrix elements
associated with higher Fock states like $|Q\overline{Q}g\!>$ are
suppressed by a factor of $v^4$ compared to the leading colour-singlet
contributions, $v$ being the average velocity of the heavy quark in
the quarkonium rest frame.\footnote{In the case of $P$-wave quarkonia,
colour-singlet and colour-octet mechanisms contribute at the same order
in $v$ to annihilation rates and production cross sections and must
therefore both be included for a consistent calculation \cite{BBL92}.}
The NRQCD description of $S$-wave quarkonia production or annihilation
thus reduces to the colour-singlet model in the non-relativistic limit
$v\to 0$. It has been shown in the rigorous analysis of Ref.\cite{BBL95}
that the factorization assumption of the CS model is correct for any
specific $S$-wave process in the non-relativistic limit to all orders
in $\alps$.

Colour-octet matrix elements can become important if their associated
short distance coefficient is enhanced compared to the colour-singlet
contribution. It has recently been demonstrated that large $p_\perp$
charmonium production in hadronic collisions can be accounted for in a
satisfactory way by including both fragmentation mechanisms as well as
higher Fock state contributions \cite{BF95,COM_TEVATRON}. For the
colour-octet matrix elements associated with the higher Fock states no
simple relation exists in general between decay and production matrix
elements. The corresponding contributions thus involve unknown
non-perturbative parameters which, in the analyses
Refs.\cite{BF95,COM_TEVATRON}, have been fitted to the experimental
data.  Once they have been measured in some production process they
can be used to predict cross sections for different energies and
different beam types. It should however be kept in mind that the
analyses carried out so far are leading-order analyses. They are
therefore plagued by large scale dependences and have to rely on the
assumption that the perturbative expansion of the short-distance
coefficients is well behaved and that the higher-order corrections do
not strongly depend on the specific production mechanism and the
collision energy.  Moreover, it has been argued recently that
important higher-twist effects have to be included in the theoretical
description of charmonium hadroproduction \cite{HT_HAD}. These effects
can yet not be predicted from first principles.

Many channels contribute to the generation of $J/\psi$ particles in
photoproduction experiments \cite{JST92}, similarly to the case of
hadroproduction. Theoretical interest so far has focussed on two
mechanisms for $J/\psi$ photo- and electroproduction,
elastic/diffractive \cite{ELASTIC,ELASTIC_EXP} and inelastic
production through photon-gluon-fusion \cite{BJ81}. While one expects
to shed light on the physical nature of the pomeron by the first
mechanism, inelastic $J/\psi$ production provides information on the
distribution of gluons in the nucleon \cite{MNS87}. The two mechanisms
can be separated by measuring the $J/\psi$ energy spectrum, described
by the scaling variable
\beq\label{ZDEF}
z = {p\cdot k_\psi}\, / \, {p\cdot k_\gamma}
\eeq
with $p, k_{\psi,\gamma}$ being the momenta of the nucleon and
$J/\psi$, $\gamma$ particles, respectively. In the nucleon rest frame,
$z$ is the ratio of the $J/\psi$ to the $\gamma$ energy,
$z=E_\psi/E_\gamma$.  For elastic/diffractive events $z$ is close to
one; a clean sample of inelastic events can be obtained in the range
$z \;\simlt\; 0.9$ \cite{BRUG}. Reducible background mechanisms, such
as $B\overline{B}$ production with subsequent decay into $J/\psi$
particles or resolved photon processes might become sizable at HERA
energies but can be easily eliminated by applying suitable cuts
\cite{JST92}.

In contrast to the case of hadroproduction, higher Fock state
contributions to inelastic $J/\psi$ photoproduction are strongly
suppressed compared to the leading colour-singlet mechanism.
Colour-octet contributions can become important only in the elastic
domain, $z\approx 1$, where the short distance coefficient of the
colour-singlet amplitude is suppressed by a factor $\alps$.  In the
inelastic region, $z\;\simlt\; 0.9$, a perturbative gluon has to be
emitted in the final state for kinematical reasons. This gluon carries
off colour charge so that a colour-singlet state can be produced
without any additional short-distance suppression. Since the
non-perturbative matrix element associated with the colour-singlet
state is enhanced compared to the colour-octet matrix element by a
factor $1/v^4$, higher Fock state contributions to inelastic $J/\psi$
photoproduction can safely be neglected. Fragmentation mechanisms
dominate the production of $J/\psi$ particles at large $p_\perp$
\cite{PT_TEV} but their contribution to the total cross section is
very small.  As a result, the dominant mechanism for inelastic
$J/\psi$ photoproduction is by the colour-singlet state (as assumed in
the CS model) and the cross section can be predicted without including
unknown non-perturbative parameters.

A comparison of the leading-order predictions with photoproduction
data of fixed-target experiments \cite{NA14,FTPS} reveals that the
$J/\psi$ energy dependence $\md\sigma/\md{}z$ is adequately accounted
for in the inelastic region $z\;\simlt\; 0.9$.  The theoretical
calculation however underestimates the normalization of the measured
cross section in general by more than a factor two, depending in
detail on the $J/\psi$ energy and the choice of the parameters
\cite{JST92}. The discrepancy with cross sections extrapolated from
electroproduction data \cite{EMC,NMC} is even larger.

The lowest-order approach to inelastic $J/\psi$ photoproduction
demands several theoretical refinements: (i) Relativistic corrections
due to the motion of the charm quarks in the $J/\psi$ bound state;
(ii) Higher-order perturbative QCD corrections; and last but not
least, (iii) Higher-twist effects which are not strongly suppressed
due to the fairly low charm-quark mass. While the problem of
higher-twist contributions has not been quantitatively approached so
far (see Ref.\cite{HT_PHOT} for a recent discussion), the relativistic
corrections have been demonstrated to be under control in the
inelastic region \cite{JKGW93,KM83}. Including higher-order QCD
corrections is, however, expected to be essential. Their contribution
in general not only changes the overall normalization of the cross
section but can also affect the shape of inclusive differential
distributions.  Expected {\em a priori} and verified subsequently, the
QCD corrections dominate the relativistic corrections in the inelastic
region, being of the order of several $\alps(M_{J/\psi}^{2})\sim 0.3$.
In the first step of a systematic expansion, they can therefore be
determined in the static approach \cite{BBL95}.

In this work we present the complete calculation of the higher-order
perturbative QCD corrections to inelastic $J/\psi$ photoproduction;
first results have already been published in Refs.\citer{PHD,MK95}.
We perform a comprehensive analysis of total cross sections and
differential distributions for the energy range of the fixed-target
experiments and for inelastic $J/\psi$ photoproduction at HERA. A
comparison of the next-to-leading order prediction with the
experimental data will not only provide information on the gluon
distribution of the proton but appears to be a clean test for the
underlying picture of quarkonium production as developed in the
perturbative QCD sector.

\section{The Born cross section}\label{SEC_JPBORN}
\begin{figure}[hbtp]

\hspace*{1.4cm}
\epsfig{%
file=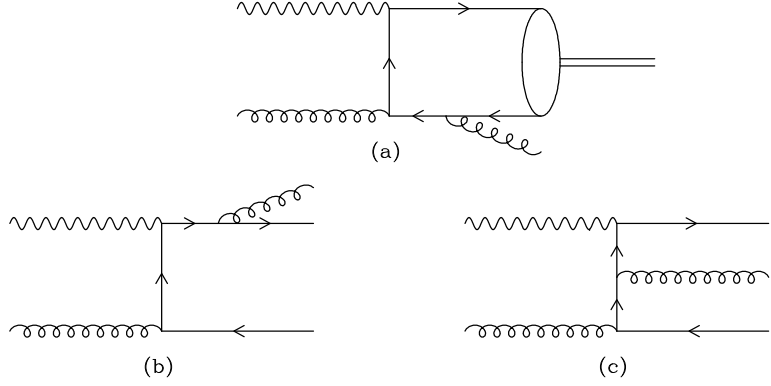,%
height=9cm,%
width=12.5cm,%
bbllx=1.0cm,%
bblly=1.9cm,%
bburx=19.4cm,%
bbury=26.7cm,%
rheight=8.2cm,%
rwidth=15cm,%
angle=-90}

\vspace*{-2.85cm}

\caption[ggdiag]{\label{FIG_JPBORNFULL}
                 The leading-order Feynman diagrams contributing
                 to inelastic $J/\psi$ photoproduction. Additional
                 graphs are obtained by reversing the arrows on the
                 heavy quark lines.}

\end{figure}

Inelastic $J/\psi$ photoproduction through photon-gluon fusion is
described in leading order by the subprocess
\begin{equation}\label{EQ_SUBPR}
\gamma(k_1) + g(k_2) \to J/\psi(P) + g(k_3)
\end{equation}
as shown in Fig.\ref{FIG_JPBORNFULL}.
%-----------------------------------------------------------------------------
%------------------------------------------------------------------------
Colour conservation and the Landau-Yang theorem \cite{LY} require the
emission of a gluon in the final state. The cross section is generally
calculated in the static approximation in which the motion of the
charm quarks in the bound state is neglected. In this approximation
the production amplitude factorizes into the short distance amplitude
${\cal{M}}(\gamma+g\to c\overline{c}+g)$, with $c\bar{c}$ in the
colour-singlet state and zero relative velocity of the quarks, and the
wave function $\varphi(0)$ of the $J/\psi$ bound state at the origin:
\beq\label{EQ_JPFIN}
{\cal{M}}(\gamma+g\to J/\psi + g) = \sqrt{\frac{2}{M_{J/\psi}}}\;
\varphi(0)\;\mbox{Tr}\Big\{\Pi_{S=1,s_z}(P_{J/\psi},M_{J/\psi})\;
{\cal{M}}(\gamma+g\to c\overline{c}+g) \Big\}  \quad .
\eeq
The spin projection operator $\Pi$ combines the quark and antiquark
spins to the appropriate triplet states. For negligible binding energy
or, equivalently, for $m_c = M_{J/\psi}/2$ one obtains \cite{GKPR80}
\beq\label{EQ_PROJOP}
\Pi_{S=1,s_z}(P_{J/\psi},M_{J/\psi}) =
\frac{1}{\sqrt{2}}\,\varepsilon\hspace*{-0.17cm}/(s_z)
\frac{P_{J/\psi}\hspace*{-0.7cm}/\hspace*{0.5cm}+M_{J/\psi}}{2}
\quad ,
\eeq
where $\varepsilon^\mu(s_z)$ is the $J/\psi$ polarization vector.
The coupling strength of the $J/\psi$ to the $c\bar{c}$-pair is
specified in terms of the orbital wave function at the origin in
momentum space. In leading order, the wave function is related to
the leptonic width according to
\beq
\Gamma_{ee} = \frac{16\pi\alpha^2e_c^2}{M_{J/\psi}^{2}}\, |\varphi(0)|^2
\quad .
\eeq
We will describe the cross section in terms of the kinematical
(pseudo-) Mandelstam variables
\bea\label{INVDEF1}
 s_1 &\!\! \equiv &\!\! s - M_{J/\psi}^{2} = (k_1+k_2)^2 - M_{J/\psi}^{2}
                                                     \nonumber \\
 t_1 &\!\! \equiv &\!\! t - M_{J/\psi}^{2} = (k_1-P)^2   - M_{J/\psi}^{2}
                                                     \nonumber \\
 u_1 &\!\! \equiv &\!\! u - M_{J/\psi}^{2} = (k_2-P)^2   - M_{J/\psi}^{2}
\eea
where $s + t + u = M_{J/\psi}^{2}$. All incoming and outgoing
particles are taken to be on-mass-shell, $k_1^2 = k_2^2 = k_3^2 = 0$
and $P^2=M_{J/\psi}^{2}$. In the static approximation
each of the heavy quarks carries one half the mass and one half
the four-momentum of the $J/\psi$:
\beq
p_c = p_{\overline{c}} = P/2 \equiv p\; ;\quad\quad
m_c = M_{J/\psi}/2\; ;\quad \mbox{and} \quad
p^2 = m_{c}^{2} \equiv m^2
\quad .
\eeq
{}From (\ref{EQ_JPFIN}) we find for the amplitude of
Fig.\ref{FIG_JPBORNFULL}(a)
\bea\label{EQ_MEBJ}
{\cal{M}} & = & \sqrt{\frac{2}{M_{J/\psi}}}\;\varphi(0)\;
\frac{1}{2\sqrt{3}}\delta_{ab} \;\, g^2 e e_c \;
\varepsilon^{\mu}(k_1)
\varepsilon^{\nu}(k_2)
\varepsilon^{\lambda}(k_3)  \nonumber \\[1mm]
& &\quad\times
\frac{4}{\sqrt{2}} \; \mbox{Tr} \left\{
\varepsilon\hspace*{-0.17cm}/_{J/\psi}(s_z)
(p\hspace*{-0.17cm}/ + m)\gamma_{\mu}
\frac{p\hspace*{-0.17cm}/ - k\hspace*{-0.17cm}/_1 + m}{t_1}
\gamma_{\nu}
\frac{-p\hspace*{-0.17cm}/ - k\hspace*{-0.17cm}/_3 + m}{s_1}
\gamma_{\lambda}\right\}
\quad .
\eea
Here $g$ and $e$ are the strong and electromagnetic couplings
respectively, $g=\sqrt{4\pi\alps}$, $e=\sqrt{4\pi\alpha}$, and
$e_c$ is the magnitude of the charm quark charge in units of $e$.
Charge conjugation invariance implies that the Feynman graphs are
symmetric under reversion of the fermion flow. All six amplitudes
contributing to the leading-order cross section are proportional to
the same colour factor $1/\sqrt{3}\;\delta_{ij}\;(T^aT^b)_{ij}$ =
$1/(2\sqrt{3})\;\delta_{ab}$, where $a$ and $b$ are the colour indices
of the gluonic quanta. Gauge invariance ensures that we may
sum over the spins of the initial photon and over the gluon spins by
employing the substitutions $\sum\varepsilon_\mu\varepsilon_\nu
= - g_{\mu\nu}$. To sum over the $J/\psi$ spin states, we use $
\sum
\varepsilon_{J/\psi}^{\rho}
\varepsilon_{J/\psi}^{\sigma} =
- g^{\rho\sigma} + P^{\rho}P^{\sigma}/M^{2}_{J/\psi}
$.
Averaging over the initial photon/gluon helicities and colour the
result for the cross section of the subprocess (\ref{EQ_SUBPR})
may be written as \cite{BJ81}
\beq\label{EQ_JPSIGB4DIM}
\frac{\md\sigma^{(0)}}{\md t_1} =
\frac{128\pi^2}{3}\, \frac{\alpha\alpsq e_c^2}{s^2}\,
 M^{2}_{J/\psi}\, \frac{|\varphi(0)|^2}{M_{J/\psi}}\;
\frac{s^2\,s_1^2 + t^2\,t_1^2 + u^2\,u_1^2}{s_1^2\;t_1^2\;u_1^2}
\quad .
\eeq

Although the cross section (\ref{EQ_JPSIGB4DIM}) is infrared finite it
is not clear {\em a priori} in which kinematical region perturbative
QCD can reliably be applied to $J/\psi$ photoproduction. Indeed, at
large $z$ the $J/\psi$ scatters more and more elastically and multiple
soft gluon emission has to be considered. Similar effects might become
important in the region where the transverse momentum $p_\perp$ of the
$J/\psi$ tends towards zero. As discussed before, it is mandatory to
require $z \;\simlt\; 0.9$ in order to eliminate contributions from
elastic/diffractive production mechanisms. In Ref.\cite{MNS87} it has
been argued that an additional cut on the transverse momentum of the
$J/\psi$, $p_\perp\;\simgt\;1$~GeV, has to be applied to define the
truly inelastic region. This conclusion was, however, based on a
comparison of leading-order cross sections with experimental data that
have later been found to be contaminated from coherent production in
the small $p_\perp$ domain \cite{EMC}. Accordingly, the region of
applicability of perturbative QCD had not been clarified completely so
far. As will become clear in Sec.\ref{SEC5}, the analysis of the
next-to-leading order corrections restricts the kinematical
domain where fixed-order calculations give a reliable description of
inelastic $J/\psi$ photoproduction. We will observe that the
perturbative QCD calculation is not under proper control in the
singular boundary region $z\to 1$ and $p_\perp \to 0$, thereby
indicating where multiple soft gluon emission becomes important.

\section{The NLO cross section}\label{SEC_3}
Including higher-order QCD corrections to the partonic reaction
(\ref{EQ_SUBPR}) is expected to be essential for the theoretical
description of inelastic $J/\psi$ photoproduction.  Next-to-leading
order corrections to open heavy flavour photoproduction have been
calculated over the recent years \cite{EN89,SV92}. They have been
found to increase the normalization of the cross section significantly
without strongly affecting the shape of the inclusive differential
distributions.  These results can however not directly be transferred
to the case of bound state production as will become clear in
Sec.\ref{SEC5}. It is thus important to investigate how the features
of the lowest-order $J/\psi$ photoproduction cross section are
modified by radiative corrections and by including new production
mechanisms which contribute in next-to-leading order.

In this work we present a complete calculation of the higher-order
perturbative QCD corrections to inelastic $J/\psi$ photoproduction.
Results for total cross sections and the $J/\psi$ energy and
transverse momentum spectrum will be discussed. The photon-parton
reactions which contribute to the inclusive cross sections up to order
$\alpha\alptr$ are
\begin{displaymath}
\begin{array}{lcll}
\hspace*{5.cm}\gamma + g & \to & (Q\overline{Q}) + g & \quad\quad
{\cal{O}}(\alpha\alpsq),\; {\cal{O}}(\alpha\alptr) \\
\hspace*{5.cm}\gamma + g  & \to & (Q\overline{Q}) + g + g & \quad\quad
{\cal{O}}(\alpha\alptr) \\
\hspace*{5.cm}\gamma + g  & \to & (Q\overline{Q}) + q + \bar{q} & \quad\quad
{\cal{O}}(\alpha\alptr) \\
\hspace*{5.cm}\gamma + q(\bar{q}) & \to & (Q\overline{Q}) + g + q(\bar{q}) &
\quad\quad {\cal{O}}(\alpha\alptr)
\end{array}
\end{displaymath}
including virtual corrections to the leading-order process.  We choose
a renormalization and factorization scheme in which the massive
particles are decoupled smoothly for momenta smaller than the heavy
quark mass \cite{MSBAR_EXT}. This implies that the heavy quark does
not contribute to the evolution of the QCD coupling and of the
structure functions.  Furthermore, there are no contributing
subprocesses initiated by an intrinsic heavy flavour coming directly
from the structure function of the proton. All effects of the heavy
quark are contained in the parton cross section.

The calculation of the the next-to-leading order corrections will be
outlined in this section. More details can be found in the
Appendices.

\subsection{Virtual corrections}
The evaluation of the ${\cal O}(\alps)$ corrections to inelastic
$J/\psi$ photoproduction involves the calculation of the virtual cross
section obtained from the interference term between the virtual and
the Born amplitude. For the unrenormalized virtual cross section one
finds in $n\equiv 4 - 2\epsilon$ dimensions
\bea\label{SIGVI}
\left[s^2\,\frac{\md^2\sigma^{(1)}}{\md t_1\md u_1}\right]^V & = &
\frac{1}{(N^2-1)}\frac{1}{4(1-\epsilon)^2}
\frac{\pi (4\pi)^{-2+\epsilon}}{\Gamma(1-\epsilon)}
\left(\frac{t_1u_1-4m^2s}{\mu^2s}\right)^{-\epsilon} \nonumber\\[1mm]
&&\hspace{1cm}\times
\delta(s_1+t_1+u_1+8m^2) \;
\sum 2\,\mbox{Re}({\cal{M}}^B{\cal{M}}^{V*})
\quad ,
\eea
where $1/(N^2-1)$ is the colour average factor for the gluon in the
initial state and $N$ denotes the number of colours. Photons and
gluons have $n-2$ spin degrees of freedom resulting in the spin
average factor $1/(n-2)^2 =1/4(1-\epsilon)^2$.  The parameter $\mu$
has the dimensions of a mass and is introduced in order to compensate
for the dimensionality of the gauge coupling constants in $n$
dimensions.

The Feynman gauge has been adopted to evaluate the 105 diagrams which
contribute to the virtual amplitude. Charge conjugation invariance
implies that the graphs are symmetric under reversion of the fermion
flow. The ultraviolet (UV), infrared (IR), and the collinear or mass
(M) singularities have been treated by using $n$ dimensional
regularization and show up as single and double pole terms of the type
$\epsilon^{-i}(i=1,2)$. We have refrained from adopting the seemingly
simpler scheme of dimensional reduction which however gives rise to
complications in the massive quark case as pointed out in
Ref.\cite{BKVS89}. The Feynman integrals containing loop momenta in
the numerator have been reduced to a set of scalar integrals using an
adapted version of the reduction program outlined in
Ref.\cite{PV79}. This program has been extended to treat $n$
dimensional tensor integrals with linear dependent propagators in
order to account for the IR- and M-singularities and the special
kinematical situation which is encountered in the nonrelativistic
approximation to $J/\psi$ photoproduction. The scalar integrals have
been calculated by the Feynman parametrization technique and
analytical results are listed in Appendix~\ref{APP1}.

\begin{figure}[hbtp]

\hspace*{1.4cm}
\epsfig{%
file=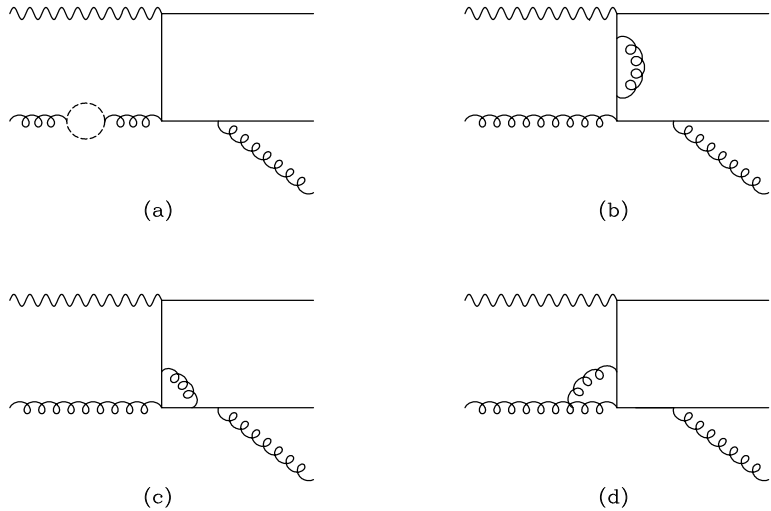,%
height=9cm,%
width=12.5cm,%
bbllx=1.0cm,%
bblly=1.9cm,%
bburx=19.4cm,%
bbury=26.7cm,%
rheight=8.2cm,%
rwidth=15cm,%
angle=-90}

\vspace*{-1.2cm}

\caption[ggdiag]{\label{SEVE}
                 Generic Feynman diagrams contributing to the
                 virtual amplitude (part~1):
                 self-energy corrections (a,b) [36~diagrams],
                 abelian vertex corrections ($\gamma c\overline{c}$-
                 and $g c\overline{c}$-vertices)~(c) [18~diagrams],
                 and non-abelian vertex corrections~(d)
                 [12~diagrams]. }
\end{figure}

The UV-divergences which are contained in the fermion self energy
graphs and the vertex corrections shown in Fig.\ref{SEVE}, are removed
by renormalization of the heavy quark mass and the QCD coupling.
%-----------------------------------------------------------------
%-----------------------------------------------------------------
The masses of the light quarks appearing in the fermion loops in
Fig.\ref{SEVE}a have been neglected while the mass parameter of the
heavy quark has been defined in the on-mass-shell scheme. The
renormalization of the QCD coupling has been carried out in the
extended $\overline{\mbox{MS}}$ scheme introduced in
Ref.\cite{MSBAR_EXT} and adopted in previous calculations of open
heavy flavour production \cite{NDE88,EN89,SV92}. In this scheme
the effects of heavy flavours are smoothly decoupled for momenta
much smaller than the heavy quark mass. The bare coupling has to be
replaced according to
\bea\label{EQ_MSBARMOD}
g_{\mbox{\scriptsize bare}}
 & = & Z_g\, g(\mursq)\nonumber \\
 & = &       g(\mursq)
\left[1 - \frac{\alps(\mursq)}{8\pi}\,
\left\{\left(\frac{1}{\epsilon}-\gamma_E+\ln{4\pi}
-\ln\left(\frac{\mursq}{\mu^2}\right)\right)
\beta_0+\frac{2}{3}\ln\left(\frac{m^2}{\mursq}
\right)\right\}\right] \quad,
\eea
with $\beta_0=(11N - 2\nf)/3$. The total number of flavours including
the heavy quark is denoted by $\nf$ and $\nlf = \nf - 1$ is the number
of light quarks. The renormalization constant $Z_g$ is defined such
that the contribution of the heavy-fermion loop in the gluon self
energy graphs is cancelled in the renormalized cross section
for small momenta flowing into the heavy-fermion loop. The
UV-divergences arising from gluon or light fermion loops are removed
according to the standard $\overline{\mbox{MS}}$ subtraction scheme
\cite{MSBAR}.  For the renormalization scale evolution of the strong
coupling one obtains from (\ref{EQ_MSBARMOD})
\beq
\frac{\partial g^2}{\partial\ln\mursq}
= g\,\beta(g)
= -\alpsq(\mursq)\left(\beta_0 + \frac{2}{3}\right)
= -\alpsq(\mursq)\beta_0(\nlf)\quad .
\eeq
The extended $\overline{\mbox{MS}}$ scheme thus implies that the heavy
quark does not contribute to the evolution of the QCD coupling so that
$\alps$ has to be evaluated using $\nlf$ active flavours.

\begin{figure}[hbtp]

\hspace*{1.4cm}
\epsfig{%
file=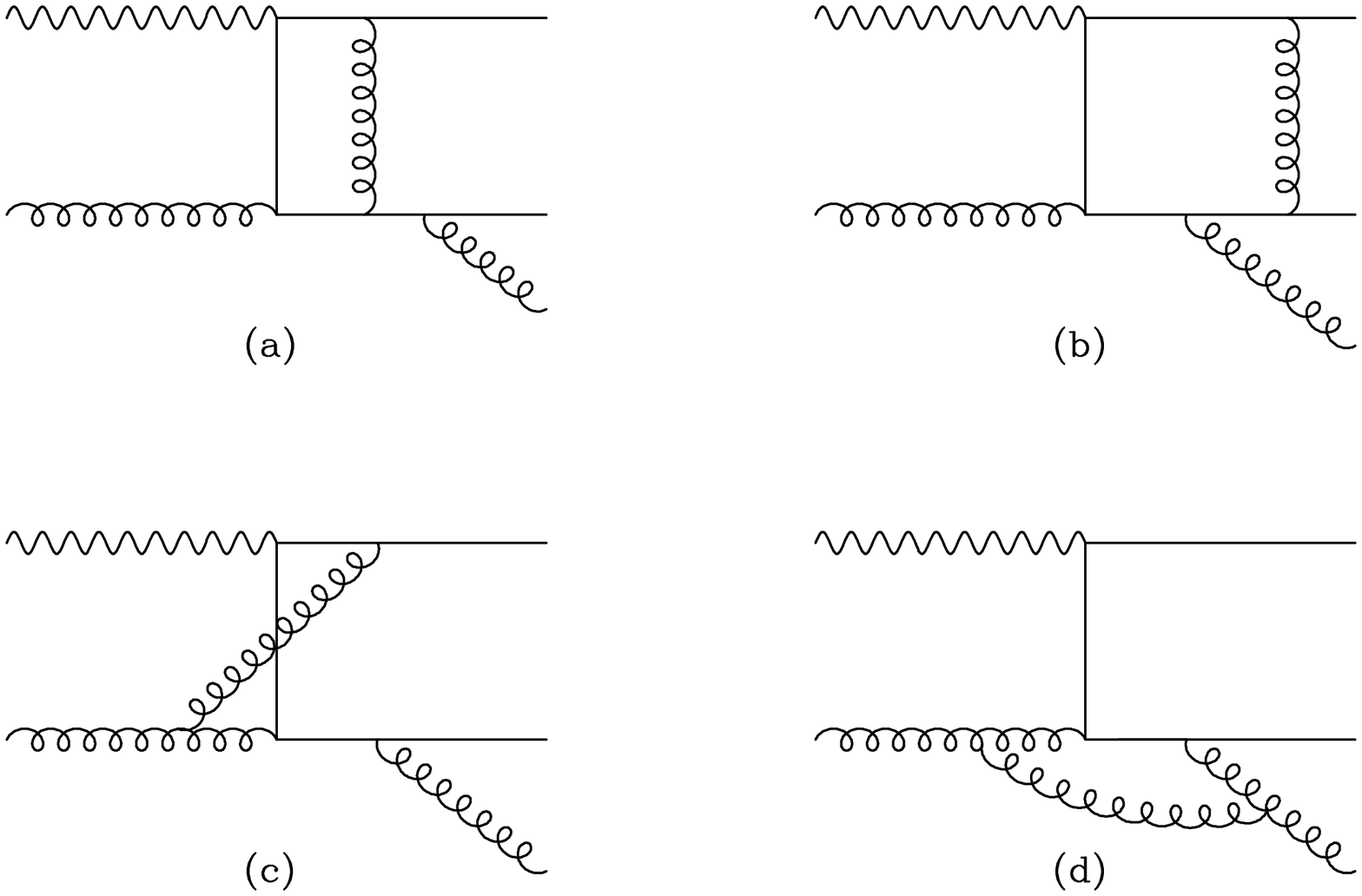,%
height=9cm,%
width=12.5cm,%
bbllx=1.0cm,%
bblly=1.9cm,%
bburx=19.4cm,%
bbury=26.7cm,%
rheight=8.2cm,%
rwidth=15cm,%
angle=-90}

\vspace*{-0.7cm}

\hspace*{1.4cm}
\epsfig{%
file=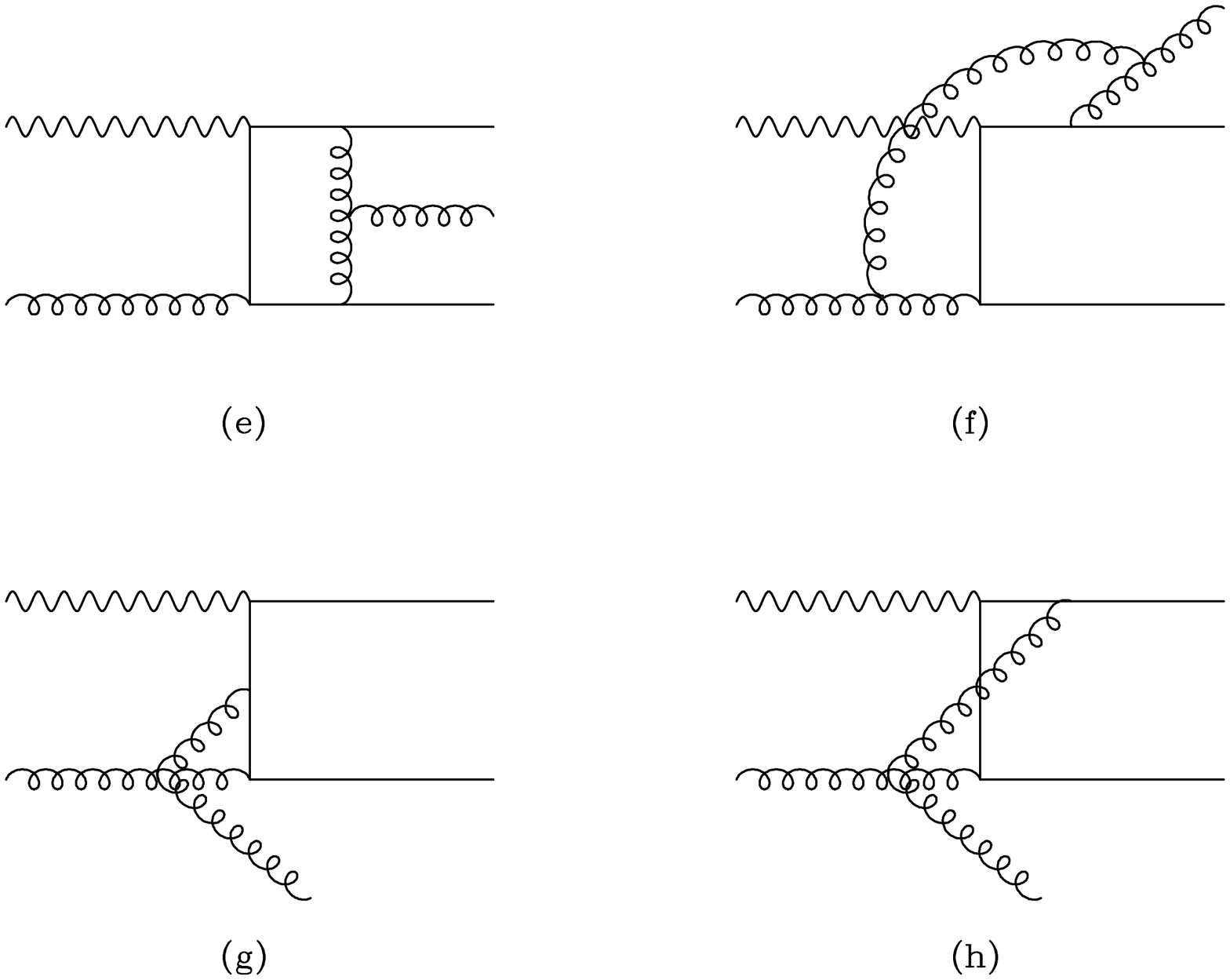,%
height=9cm,%
width=12.5cm,%
bbllx=1.0cm,%
bblly=1.9cm,%
bburx=19.4cm,%
bbury=26.7cm,%
rheight=8.2cm,%
rwidth=15cm,%
angle=-90}

\vspace*{0.8cm}

\caption[ggdiag]{\label{BOXES}
                 Generic Feynman diagrams contributing to the
                 virtual amplitude (part~2):
                 abelian four-point box graphs (a) [12~diagrams],
                 abelian five-point box graphs (b) [6~diagrams],
                 non-abelian four-point box graphs (c) [8~diagrams]
                 and (d) [4~diagrams],
                 non-abelian five-point box graphs (e) [4~diagrams]
                 and (f) [2~diagrams], and diagrams involving
                 four-gluon couplings (g) [2~diagrams] and
                 (h) [1~diagram].}
\end{figure}

Besides the self-energy diagrams and vertex corrections, 39 box graphs
contribute to the virtual amplitude, which can be grouped in eight
classes as shown in Fig.\ref{BOXES}.
%-----------------------------------------------------------------
%-------------------------------------------------------------------------
All diagrams falling into one particular class are related by exchange
of photon or gluon momenta, adjustment of the colour factor and
reversion of the fermion flow. These relations have been checked
explicitly.

The exchange of longitudinal gluons between the massive quarks
in diagram Fig.\ref{BOXES}b leads to the Coulombic singularity
$\sim \pi^2/v$ which can be isolated by introducing a small relative
quark velocity $v$ (see Appendix~\ref{APP1} for details). For the
Coulomb-singular part of the virtual cross section we find
\beq
\sigma = |\varphi(0)|^2\;\hat\sigma^{(0)}\left(1+\frac{\alps}{\pi}\,
C_{F}\,\frac{\pi^2}{v}+
\frac{\alps}{\pi}\hat{C}+{\cal{O}}(\alpsq)\right) \quad ,
\eeq
where the colour factor is given by $C_{F} = (N^2-1)/(2N)$.  The final
state interaction in the colour-singlet channel is attractive and has
to be interpreted as the Sommerfeld rescattering correction
\cite{SOMM} which can be associated with the inter-quark potential of
the bound state. The Coulomb-singular part of the virtual cross
section is universal appearing in the next-to-leading order
corrections to all production and decay processes involving $S$-wave
quarkonia.  Following the standard path \cite{HB57}, the corresponding
contribution has to be factored out and mapped into the
$c\overline{c}$ wave function:
\beq
\sigma = |\varphi(0)|^2\left(1+
\frac{\alps}{\pi}\,C_F\,\frac{\pi^2}{v}\right)
\;\hat\sigma^{(0)}\left[1+\frac{\alps}{\pi}\hat{C}+
{\cal{O}}(\alpsq)\right]
\Rightarrow |\varphi(0)|^2\;\hat
\sigma^{(0)}\left[1+\frac{\alps}{\pi}\hat{C}+
{\cal{O}}(\alpsq)\right]
\quad .
\eeq
Only the exchange of transversal gluons contributes to the
next-to-leading order expressions for the hard parton cross section.

\subsection{Real corrections}\label{SEC_JPGLBS}
\begin{figure}[hbtp]

\vspace*{-0.3cm}
\hspace*{1.4cm}
\epsfig{%
file=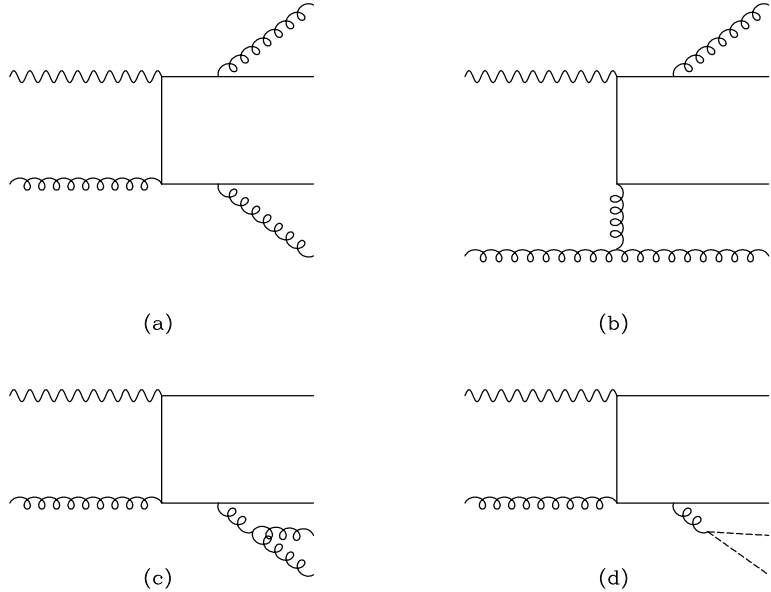,%
height=9cm,%
width=12.5cm,%
bbllx=1.0cm,%
bblly=1.9cm,%
bburx=19.4cm,%
bbury=26.7cm,%
rheight=8.2cm,%
rwidth=15cm,%
angle=-90}

\vspace*{0.35cm}

\caption[ggdiag]{\label{JPREAL}
                 Generic Feynman diagrams contributing to the
                 gluon bremsstrahlung process (a-c) and the
                 quark-antiquark splitting reaction (d).}

\vspace*{-3mm}

\end{figure}

The evaluation of the ${\cal{O}}(\alpha\alptr)$ cross section
requires the calculation of the gluon bremsstrahlung reaction
\beq\label{EQ_TWOGL}
\gamma (k_1) + g(k_2) \to J/\psi(2p) + g(k_3) + g(k_4)
\quad
\eeq
and processes where the final-state gluon splits into light
quark-antiquark pairs
\beq\label{EQ_TWOQ}
\gamma (k_1) + g(k_2) \to J/\psi(2p) + q(k_3) + \overline{q}(k_4)
\quad .
\eeq
The 48 Feynman diagrams which contribute to the amplitude can be
obtained from the generic ones shown in Fig.\ref{JPREAL} by
permutation of the photon and gluon lines. Spin and colour projection
imply that the diagrams are invariant under reversion of the fermion
flow.
%-----------------------------------------------------------------
%--------------------------------------------------------------------------
For the cross section of the two-gluon final states (\ref{EQ_TWOGL}),
averaged over initial spins and colours, one obtains (see
Appendix~\ref{APP2A})
\bea\label{JPSIGREAL}
\left[s^2\,\frac{\md^2\sigma^{(1)}}{\md t_1\md u_1}\right]^R & = &
\frac{1}{2!}\;\frac{1}{(N^2-1)}\,\frac{1}{4(1-\epsilon)^2}\;
\frac{\mu^{2\epsilon}(4\pi)^{-4+2\epsilon}}{2\Gamma(1-2\epsilon)}
\left(\frac{t_1u_1-4m^2s}{\mu^2s}\right)^{-\epsilon} \nonumber \\
& &\quad\quad\quad\quad\times
\;s_3^{-\epsilon} \;
\int d\Omega_n
\sum \left|{\cal{M}}^R\right|^2
\quad ,
\eea
where
\begin{eqnarray}\label{INVDEF}
s   &\!\! = \!\!& (k_1+k_2)^2 \equiv s_1 + 4m^2     \\ \nonumber
t_1 &\!\! = \!\!& (2p-k_1)^2 - 4m^2  \\                \nonumber
u_1 &\!\! = \!\!& (2p-k_2)^2 - 4m^2      \quad ,
\end{eqnarray}
$s_3 = (k_3 + k_4)^2 = s_1 + t_1 + u_1 + 8m^2$ and $\md\Omega_n =
\md\theta_1 \sin^{n-3} \theta_1\, \md\theta_2\sin^{n-4}\theta_2$. The
angles $\theta_1$ and $\theta_2$ which describe the orientation of the
outgoing light partons are defined in Appendix~\ref{APP2A}. A factor
$1/2!$ has to be included since there are two identical particles in
the final state. For the cross section of the
light-quark-antiquark-splitting reaction (\ref{EQ_TWOQ}) this factor
has to be dropped.

The real-gluon cross section contains IR and M singularities so that
the square of the amplitude had to be calculated in $n$ dimensions up
to order $\epsilon^2$. For the sum of the gluon polarization we have
used $\sum\varepsilon_\mu\varepsilon_\nu = - g_{\mu\nu}$ and the
unphysical longitudinal gluon polarizations have been removed by
adding ghost contributions.

The computation of the real cross section has been performed by
adopting the phase space slicing method as outlined in Ref.\cite{KP78}
and used in previous calculations of open heavy flavour production
\cite{BKVS89,SV92,LRSV94}. We have split the cross section into an
infrared-collinear part ($s_3\le\Delta$), which contains all
singularities due to soft gluon emission and splitting of the final
state gluon into gluon and light quark-antiquark pairs, and a
hard-gluon part ($s_3>\Delta$). The cut-off parameter $\Delta$ is
chosen such that it can be neglected with respect to mass
terms like $m^2$ and the kinematical invariants $s_1$, $t_1$, and
$u_1$. In the final answer the limit $\Delta \to 0$ is carried out.
The hard-gluon part ($s_3>\Delta$) contains single pole terms which
are associated with initial state gluon radiation only. These
collinear divergences have to be absorbed into the renormalization of
the parton densities as outlined in Sec.\ref{SEC_JPMFAC}. To perform
the integration over the orientation of the final state gluons or
light quarks the matrix element squared has been decomposed into sums
of terms which have at most two factors containing the dependence on
the polar angle $\theta_1$ and the azimuthal angle $\theta_2$. This
decomposition is described in Appendix~\ref{APP2A}.

The infrared-collinear cross section ($s_3\le\Delta$) for the two-gluon
final states (\ref{EQ_TWOGL}) is obtained from the expression
\bea\label{EQ_JPSOFT}
\left[s^2\,\frac{\md\sigma^{(1)}}{\md t_1\md u_1}\right]^S & = &
\frac{1}{2!}\;\frac{1}{(N^2-1)}\,\frac{1}{4(1-\epsilon)^2}\;
\frac{\mu^{2\epsilon}(4\pi)^{-4+2\epsilon}}{2\Gamma(1-2\epsilon)}
\left(\frac{t_1u_1-4m^2s}{\mu^2s}\right)^{-\epsilon}
\delta (s_1+t_1+u_1+8m^2) \nonumber\\[1mm]
&&\hspace{2cm}\times
\int_{0}^{\Delta}\md s_3 s_3^{-\epsilon}\int \md\Omega_n\sum
\left|{\cal{M}}^R\right|^2
\quad .
\eea
For the light-quark-antiquark final state the symmetry factor
$1/2!$ has to be dropped. The calculation of the infrared-collinear
cross section (\ref{EQ_JPSOFT}) is described in detail in Appendix~\ref{APP2B}.
Adding the resulting expression (\ref{EQ_MSQCOLL1},\ref{EQ_MSQCOLL2})
to the virtual correction leads to a cancellation of the infrared
singularities.

\subsection{The photon-quark subprocess}
\begin{figure}[hbtp]

\hspace*{1.4cm}
\epsfig{%
file=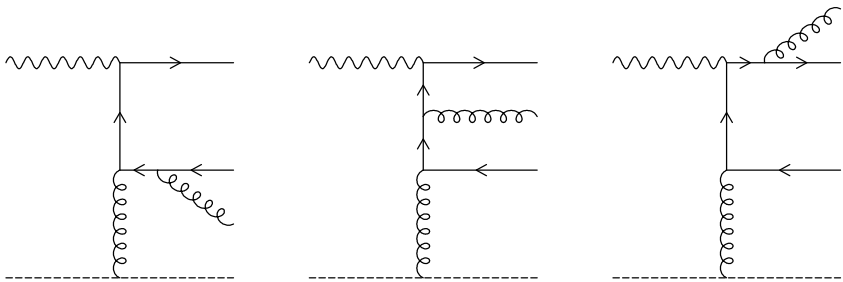,%
height=9cm,%
width=12.5cm,%
bbllx=1.0cm,%
bblly=1.9cm,%
bburx=19.4cm,%
bbury=26.7cm,%
rheight=8.2cm,%
rwidth=15cm,%
angle=-90}

\vspace*{-3.5cm}

\caption[ggdiag]{\label{JPGQ}
                 Feynman diagrams contributing to the
                 photon-(anti-)quark subprocess
                 $\gamma q(\overline{q})\to J/\psi\, gq(\overline{q})$.
                 Additional graphs are obtained by reversing the
                 arrows on the heavy quark lines.}
\end{figure}

The cross section in next-to-leading order involves a new production
mechanism where the photon is scattered off a light (anti-)quark from
the proton
\beq\label{EQ_GAMQ}
\gamma(k_1) + q(\overline{q})(k_2) \to J/\psi(2p) + g(k_3)
+ q(\overline{q})(k_4)
\quad .
\eeq
Because of spin and colour projection the $J/\psi$ particle can only
be produced in the Bethe-Heitler reaction shown in Fig.\ref{JPGQ}.
The amplitude of the $\gamma{}q(\bar{q})$ subprocess does not
depend on the electric charge of the light quarks.
%------------------------------------------------------------------------
%-----------------------------------------------------------------
Averaging over spin and colour of the initial state particles the
$\gamma{}q(\bar{q})$ cross section is given by
\beq\label{GQXSEC}
s^2\,\frac{\md^2\sigma_{q\gamma}^{(1)}}{\md t_1\md u_1}  =
\nlf\; \frac{1}{N}\frac{1}{4(1-\epsilon)}
\frac{\mu^{2\epsilon}(4\pi)^{-4+2\epsilon}}{2\Gamma(1-2\epsilon)}
\left(\frac{t_1u_1-4m^2s}{\mu^2s}\right)^{-\epsilon}\, s_3^{-\epsilon}
\int \md\Omega_n\sum\left|{\cal{M}}_{q\gamma}\right|^2 \quad .
\eeq
The cross section (\ref{GQXSEC}) contains collinear divergences $\sim
1/\epsilon$ due to gluon emission from the initial-state light
(anti-)quark which have to be removed by mass factorization.

\subsection{Mass factorization}\label{SEC_JPMFAC}
The collinear singularities contained in the hard-gluon bremsstrahlung
reaction and the $\gamma{}q(\bar{q})$ subprocess are universal and can
be absorbed, as usual, into the renormalization of the parton
densities.  According to the factorization theorem \cite{CSS86} the
parton cross section can be written as
\beq\label{EQ_FAC1}
\frac{\md^2\sigma_{\gamma{}i}(s,t_1,u_1,\mu^2,\epsilon)}{\md t_1\md u_1}
= \int_0^1 \md x\, x\;\Gamma_{gi}(x,Q^2,\mu^2,\epsilon)\;
\frac{\md ^2\hat{\sigma}_{\gamma{}g}(\hat{s},{t}_1,\hat{u}_1,Q^2)}
     {\md{t}_1\md\hat{u}_1}
\quad ,
\eeq
where $\hat{s}=xs$, $\hat{u}_1=xu_1$ and $i=g,q(\bar{q})$.  The
reduced cross sections $\md\hat\sigma_{\gamma{}g}$ and
$\md\hat\sigma_{\gamma{}q(\bar{q})}$ defined by the above equation are
free of collinear singularities and depend on the mass factorization
scale $Q^2$. This scale separates long and short distance effects and
is a priori only determined to be of the order of the heavy quark mass
$m$. The collinear singularities are contained in the splitting
functions $\Gamma_{gi}$ of the incoming partons (gluons or light
quarks). The splitting functions depend on the mass factorization
scale $Q^2$ and further on the parameter $\mu^2$ which is an artefact
of $n$-dimensional regularization. Up to leading order in $\alps$
they are given by \cite{GA79}
\beq\label{EQ_SPLITX}
\Gamma_{ij}(x,Q^2,\mu^2,\epsilon) = \delta_{ij}\delta(1-x) +
\frac{\alps}{2\pi} \left[-\frac{1}{\epsilon} P_{ij}(x)
+ f_{ij}(x,Q^2,\mu^2) \right]
\label{GAMMAIJ}
\quad ,
\eeq
where the universal Altarelli-Parisi splitting kernels \cite{AP77}
are denoted by $P_{ij}(x)$. The finite functions $f_{ij}$ are completely
arbitrary, different choices corresponding to different factorization
schemes. Here we have adopted the $\overline{\mbox{MS}}$-scheme
corresponding to
\beq\label{EQ_SPLITT1}
f_{ij}^{\overline{\mbox{\scriptsize{}MS}}}(x,Q^2,\mu^2) =
P_{ij}(x)\;\left(
\gamma_E - \ln 4\pi + \ln\frac{Q^2}{\mu^2} \right)
\quad .
\eeq
For the reduced cross section of the $\gamma$-gluon process one finds
from (\ref{EQ_FAC1})
\bea
\lefteqn{\frac{\md^2\hat{\sigma}^{(1)}_{\gamma{g}}
(s,t_1,u_1,Q^2)}{\md t_1\md u_1} =
\frac{\md ^2\sigma^{(1)}_{\gamma{g}}
(s,t_1,u_1,\mu^2,\epsilon)}{\md t_1\md u_1}}
\nonumber \\[1mm]
& & - \frac{\alps}{2\pi}
\int_0^1 \md x\, x\; P_{gg}(x)
\left[-\frac{1}{\epsilon} +\gamma_E -\ln 4\pi
+\ln\frac{Q^2}{\mu^2}\right]\;\frac{\md^2\sigma^{(0)}_{\gamma{g}}
(x s,t_1,x u_1)}{\md{t}_1\md\hat{u}_1}
\quad .
\eea
The corresponding expression for the $\gamma{}-q(\bar{q})$
scattering reaction reads
\bea\label{EQ_FAC3}
\lefteqn{\frac{\md^2\hat{\sigma}^{(1)}_{\gamma{q}}
(s,t_1,u_1,Q^2)}{\md t_1 \md u_1} = \frac{\md^2\sigma^{(1)}_{\gamma{q}}
(s,t_1,u_1,\mu^2,\epsilon)}{\md t_1\md u_1}} \nonumber \\[1mm]
& & - \frac{\alps}{2\pi}\int_0^1 \md x\, x\; P_{gq}(x)
\left[-\frac{1}{\epsilon} +\gamma_E -\ln 4\pi +\ln\frac{Q^2}{\mu^2}
\right]\;\frac{\md^2\sigma^{(0)}_{\gamma{g}}(x s,t_1,x u_1)}{\md t_1
\md\hat{u}_1} \quad .
\eea
The Altarelli-Parisi kernel $P_{gg}(x)$ of the gluon-splitting
function has the form
\bea \label{EQ_PGG}
P_{gg}(x) & = & N\left[\Theta(1-x-\delta)\cdot 2\left\{\frac{1-x}{x}
 +\frac{x}{1-x}+x(1-x)\right\}\right.\nonumber\\[1mm]
&& \hphantom{N [}\left.+\;\delta(1-x)\left(2\ln\delta+\frac{11}{6}
   \right)\vphantom{\frac{2}{1-x}}\right] -\frac13\;
   \nlf\;\delta(1-x)\quad ,
\eea
where the number of light flavours is denoted by $\nlf$.  We have
adopted the convention introduced in Ref.\cite{KP78} to regulate the
pole at $x=1$. The parameter $\delta$ allows one to distinguish
between soft ($x>1-\delta$) and hard ($x<1-\delta$) gluons and is
related to the cut-off parameter $\Delta$ by $\delta=\Delta/(s+u_1)$.
Finally, the Altarelli-Parisi kernel $P_{gq}$ appearing in
(\ref{EQ_FAC3}) is given by
\beq\label{EQ_PGQ}
P_{g\bar q}(x) = P_{gq}(x) = C_F\left[\frac{1+(1-x)^2}{x}\right]
\quad .
\eeq

After mass factorization and cancellation of the infrared
singularities between the virtual corrections and the contribution of
soft gluon emission we obtain a finite expression for the
${\cal{O}}(\alpha\alptr)$ inelastic $J/\psi$ photoproduction cross
section. An analytical result has been derived for the double
differential one-particle-inclusive cross section $\md\sigma/\md t_1
\md u_1$. The corresponding expressions are however too long to be
presented here.

\section{The parton cross section}

\begin{figure}[btp]
\hspace*{1.4cm}
\epsfig{%
file=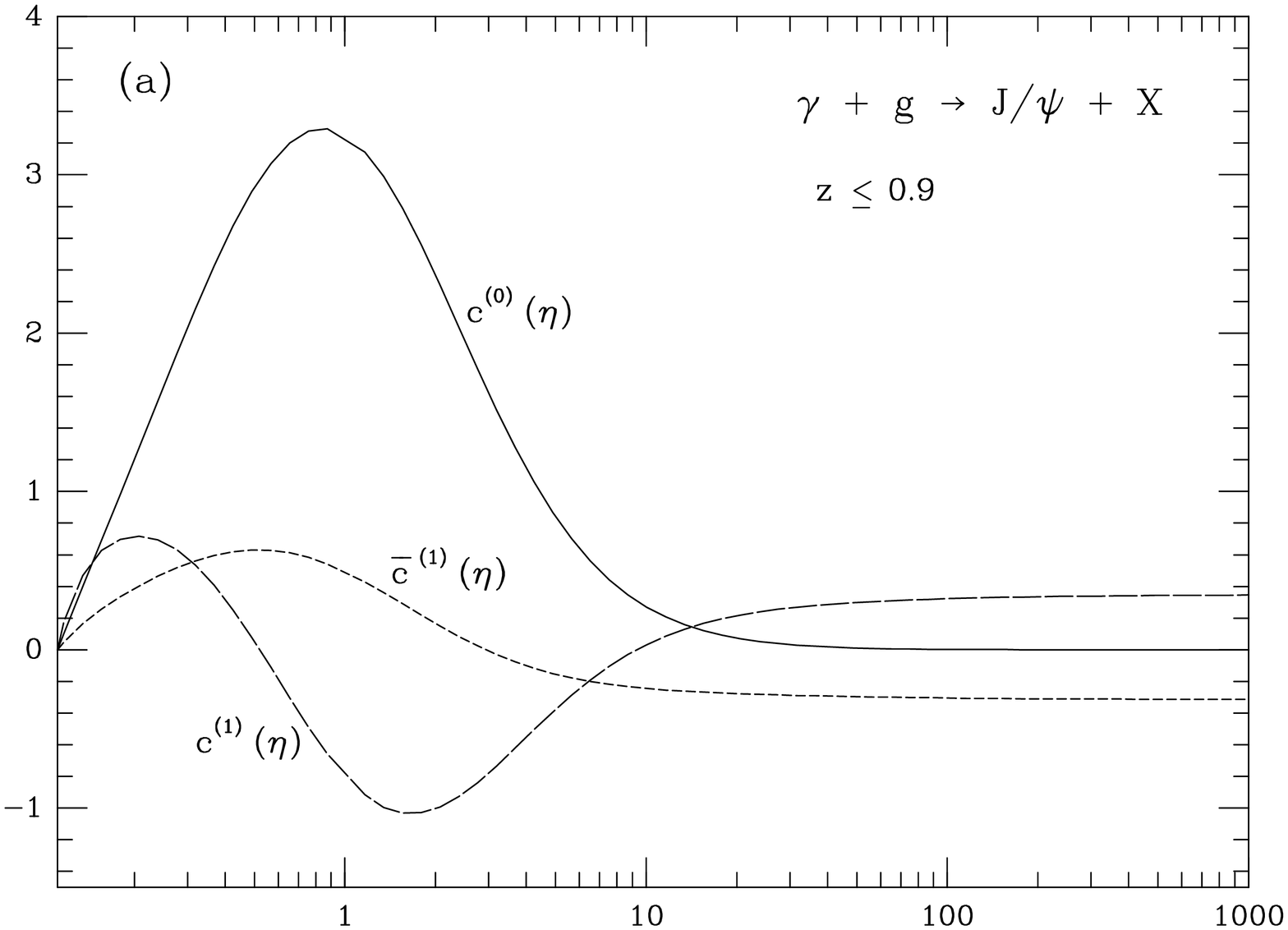,%
height=9cm,%
width=12.5cm,%
bbllx=1.0cm,%
bblly=1.9cm,%
bburx=19.4cm,%
bbury=26.7cm,%
rheight=8.2cm,%
rwidth=15cm,%
angle=-90}

\vspace*{2.0cm}

\hspace*{1.4cm}
\epsfig{%
file=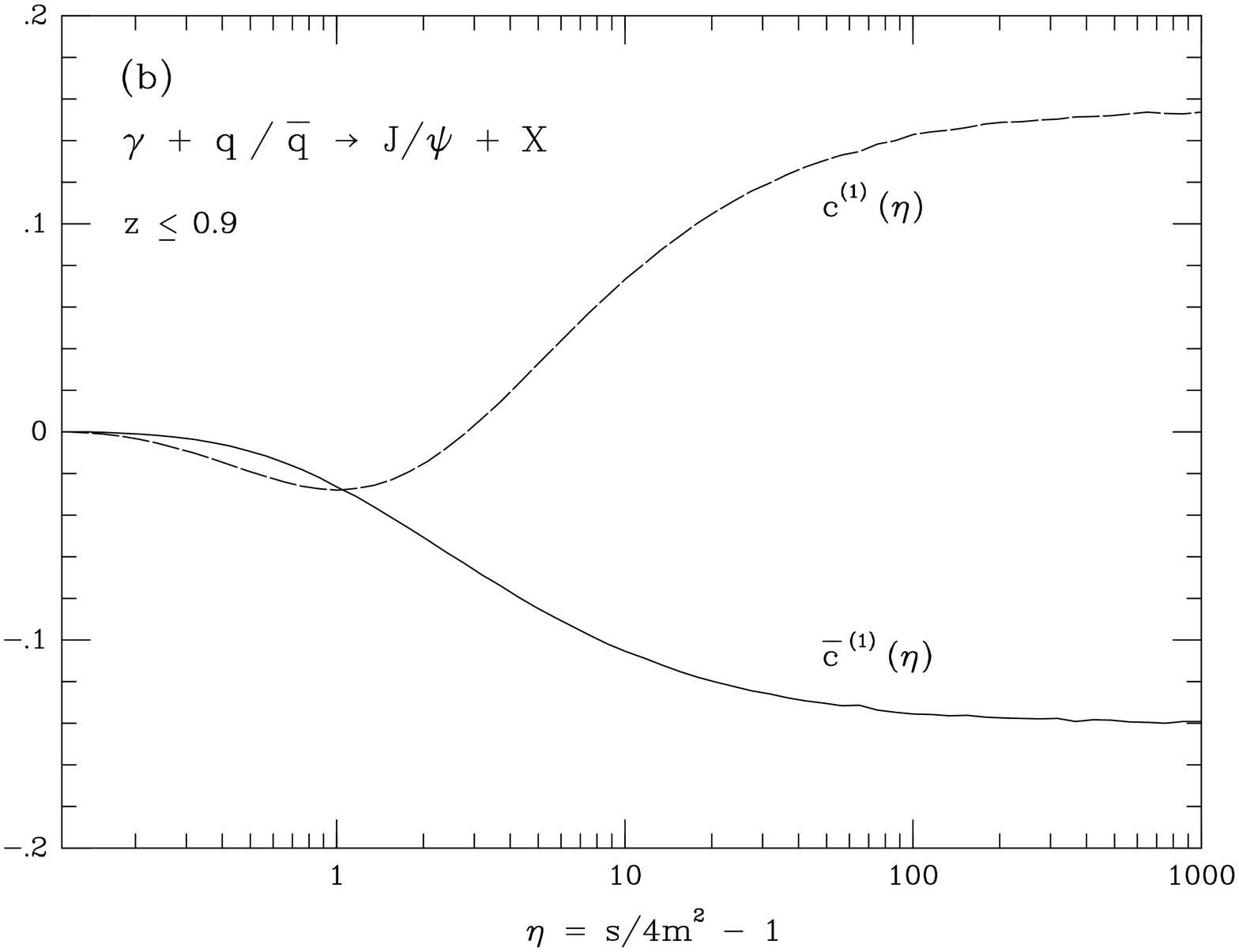,%
height=9cm,%
width=12.5cm,%
bbllx=1.0cm,%
bblly=1.9cm,%
bburx=19.4cm,%
bbury=26.7cm,%
rheight=8.2cm,%
rwidth=15cm,%
angle=-90}

\vspace*{1.0cm}

\caption[xx]{  \label{F_SCALE}
                 (a) Coefficients of the QCD corrected total inelastic
                 [$z \le 0.9$] cross section $\gamma + g \to  J/\psi + X$
                 in the physically relevant range of the scaling variable
                 $\eta = s_{\gamma p}/4m^2 - 1$; and (b) for $\gamma +
                 q/\overline{q} \to  J/\psi + X$.}

\end{figure}

The perturbative expansion of the total photon-parton cross section
can be expressed in terms of scaling functions,
\pagebreak[3]
\bea
\hat\sigma_{i\gamma}(s,m_c^2) & \!\! = & \!\!
\frac{\alpha\alpsq e^2_c}{m_c^2}\,\frac{|\varphi(0)|^2}{m_c^3}
\left[\vphantom{\ln\frac{\mursq}{Q^2}} c_{i\gamma}^{(0)}(\eta)
+ 4\pi\alps \left\{\vphantom{\ln\frac{\mursq}{Q^2}}
 c_{i\gamma}^{(1)}(\eta)
+ \overline{c}^{(1)}_{i\gamma}(\eta)\ln\frac{Q^2}{m_c^2}
\right.\right. \nonumber \\ && \hphantom{kkkkkk \!\!
\frac{\alpha\alpsq e^2_c}{m_c^2}\,\frac{|\varphi(0)|^2}{m_c^3}
[\vphantom{\ln\frac{\mursq}{Q^2}} c_{i\gamma}^{(0)}(\eta)
+ 4\pi\alps \{\vphantom{\ln\frac{\mursq}{Q^2}} }
\left.\left.
+\;\frac{\beta_0(\nlf)}{8\pi^2}\;c_{i\gamma}^{(0)}(\eta)
\ln\frac{\mursq}{Q^2}\right\}\right]\quad ,
\eea
where $i=g,q,\overline{q}$ denote the parton targets and
$\beta_0(\nlf) = (11N - 2\nlf)/3$. The scaling functions depend
on the energy variable $\eta = s/4m_c^2 - 1$. $c_{\gamma g}^{(0)}$ is
the lowest-order contribution which scales $\sim \eta^{-1}\sim
4m_c^2/s$ asymptotically. The cross section is put into a form in
which the renormalization scale $\mur$ and the factorization scale
$Q$ can be varied independently.
The scaling functions $c_{\gamma i}(\eta)$ are shown in
Figs.\ref{F_SCALE}(a) and (b) for the parton cross sections integrated
over $z\le z_1$ where we have chosen $z_1 = 0.9$ as discussed before.
%---------------------------------------------------------------------------
%---------------------------------------------------------------------------
%---------------------------------------------------------------------------
\begin{figure}[hbtp]

\hspace*{1.4cm}
\epsfig{%
file=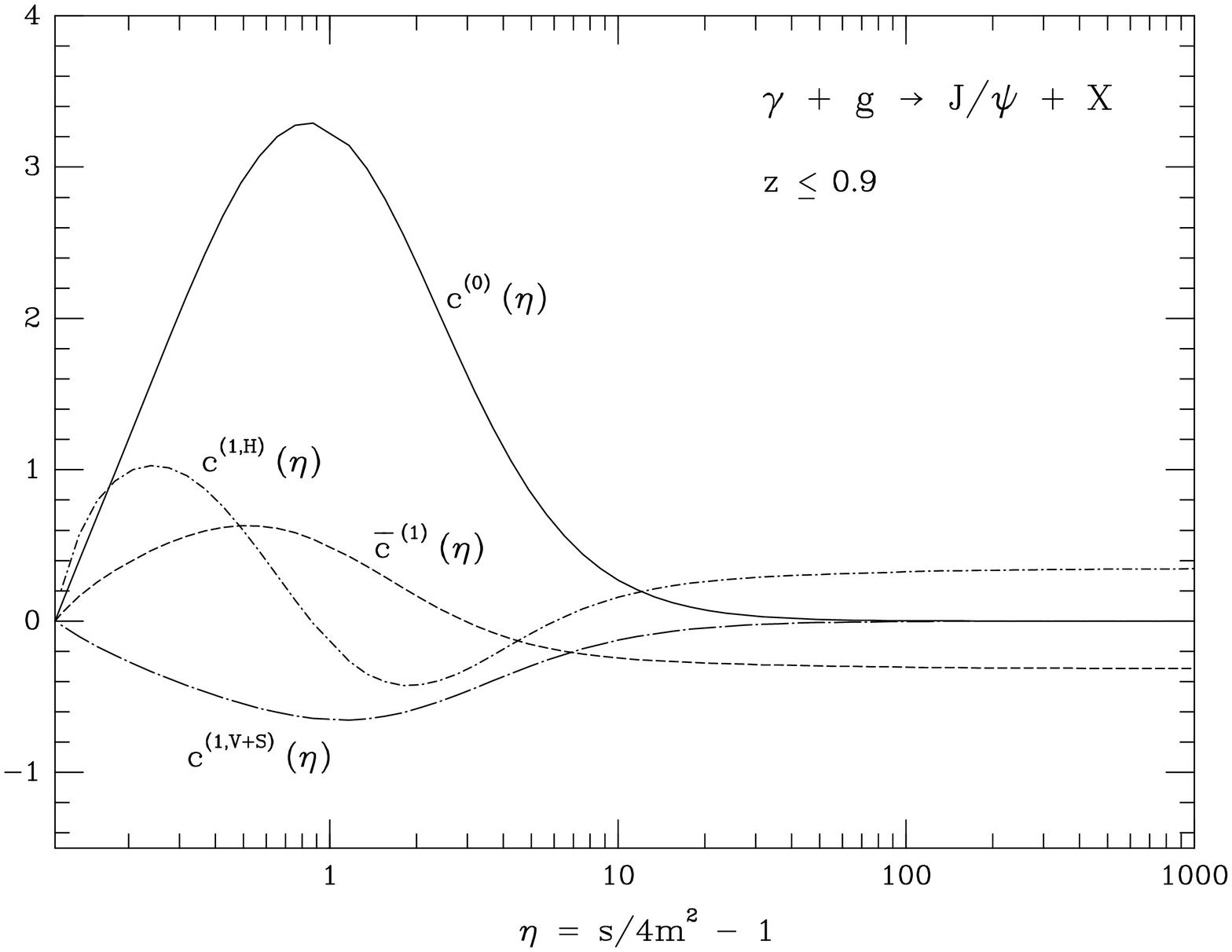,%
height=9cm,%
width=12.5cm,%
bbllx=1.0cm,%
bblly=1.9cm,%
bburx=19.4cm,%
bbury=26.7cm,%
rheight=8.2cm,%
rwidth=15cm,%
angle=-90}

\vspace*{0.5cm}

\caption[xx]{ \label{F_SCALE2}
              Coefficients of the QCD corrected total inelastic
              [$z \le 0.9$] cross section $\gamma + g \to  J/\psi + X$
              split into a hard piece and a virtual plus soft piece.}

%\vspace*{-3mm}

\end{figure}
%---------------------------------------------------------------------------
[Note that the definition of $z$ is the same at the nucleon and parton
level since the momentum fraction $x$ of the partons cancels in the
ratio $z = {p\cdot k_\psi}\, / \, {p\cdot k_\gamma}$.]  In
Fig.\ref{F_SCALE2} the scaling function of the gluon initiated parton
process has been decomposed into a "virtual + soft" (V+S) piece and a
"hard" (H) gluon-radiation piece.  The $\ln^{i}\Delta$ singularities
of the (V+S) cross sections are mapped into (H), cancelling the
equivalent logarithms in this contribution so that the limit
$\Delta\to 0$ can safely be carried out. The nomenclature "hard" and
"virtual + soft" is therefore a matter of definition, and negative
values of $c^{(\mbox{\scriptsize{}H})}$ may occur in some regions of
the parameter space.  [In the range
$0.2{\,\,}\simlt{\,\,}\eta{\,\,}\simlt{\,\,}2$ the hard
gluon-radiation piece $c_{g\gamma}^{(\mbox{\scriptsize{}1,H})}$ as
well as $\overline{c}_{\gamma g}^{(1)}$ differ from the curves in
Ref.\cite{KZSZ94} by a few percent since the experimental cut $z <
0.9$ was not implemented properly in one term of Ref.\cite{KZSZ94}.]

The following comments can be inferred from the figures.
(i) The form of the hard-gluon radiation piece
$c^{(\mbox{\scriptsize{}H})}$ resembles the corresponding scaling
function in open-charm photoproduction \cite{SV92}. The logarithmic
enhancement near threshold can be attributed to initial state gluon
bremsstrahlung.  The "virtual + soft" contribution for $J/\psi$
production is, however, significantly more negative than for
open-charm production. The destructive interference with the
lowest-order amplitude is not unplausible though, as the momentum
transfer of virtual gluons has a larger chance [in a quasi-classical
approach] to scatter quarks out of the small phase-space element
centered at $p_c + p_{\overline{c}} = p_{J/\psi}$ than to scatter them
from outside into this small element.
(ii) While $c_{g\gamma}^{(0)}$ and
$c_{g\gamma}^{(\mbox{\scriptsize{}1,V+S})}$ scale asymptotically $\sim
1/s$, the hard coefficients $c_{g\gamma}^{(\mbox{\scriptsize{}1,H})}$
and $c_{q\gamma}^{(1)}$ [as well as $\overline{c}_{g,q\gamma}^{(1)}$]
approach plateaus for high energies, built-up by the flavour excitation
mechanism.
(iii) The cross sections on the quark targets are more than one order
of magnitude smaller than those on the gluon target.
(iv) A more detailed presentation of the spectra would reveal that the
perturbative analysis is not under proper control in the limit $z\to
1$, as anticipated for this singular boundary region (see the detailed
discussion in Sec.\ref{SEC5}). Outside the diffractive region, i.e. in
the truly inelastic domain, the perturbation theory is well-behaved
however.

\section{The photon-proton cross section}\label{SEC5}
The results for inelastic $J/\psi$ production in photon-proton
collisions
\beq
\gamma + P \to J/\psi + X
\eeq
are obtained from the partonic cross sections by convolution with the
gluon and light-quark distributions $f^{P}_{i}$ in the proton
\beq
\md\sigma^{\gamma{P}} = \sum_{i=g,q(\overline{q})}\;\int\!\md{x}\,
                        f^{P}_{i}(x,Q^2)\:\md\hat\sigma^{\gamma i}
\quad .
\eeq
In the following we present a comprehensive analysis of total cross
sections and differential distributions for the energy range of the
fixed-target experiments and for inelastic $J/\psi$ photoproduction at
HERA.

\subsection{The energy range of the fixed target experiments}
\label{SEC51}
Inelastic $J/\psi$ photoproduction has been measured in fixed-target
experiments \cite{NA14,FTPS} at photon energies near $E_\gamma =
100$~GeV, corresponding to invariant energies of about
$\sqrt{s\hphantom{tk}}\!\!\!\!\! _{\gamma p}\,\, \approx 14$~GeV.
Before comparing the theoretical predictions with the experimental
data we will examine the effect of the next-to-leading order
corrections in some detail.

\begin{figure}[htbp]

\hspace*{1.4cm}
\epsfig{%
file=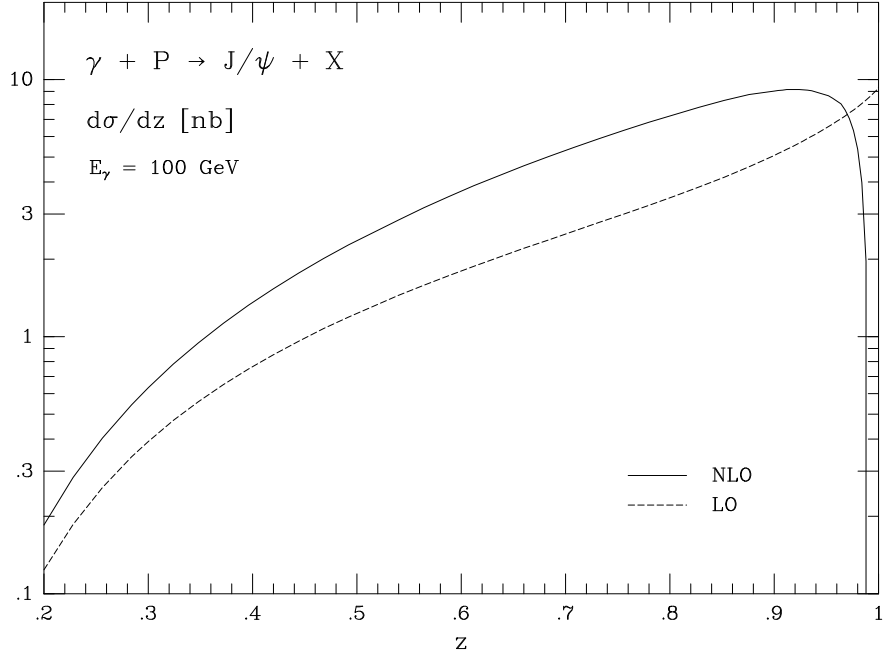,%
height=9cm,%
width=12.5cm,%
bbllx=1.0cm,%
bblly=1.9cm,%
bburx=19.4cm,%
bbury=26.7cm,%
rheight=8.2cm,%
rwidth=15cm,%
angle=-90}

\vspace*{0.5cm}

\caption[ggdiag]{\label{JPZ1}
                 Energy distribution $\md\sigma/\md{}z$
                 at an initial photon energy of $E_\gamma = 100$~GeV.}

\end{figure}

\begin{figure}[hbtp]

\hspace*{1.4cm}
\epsfig{%
file=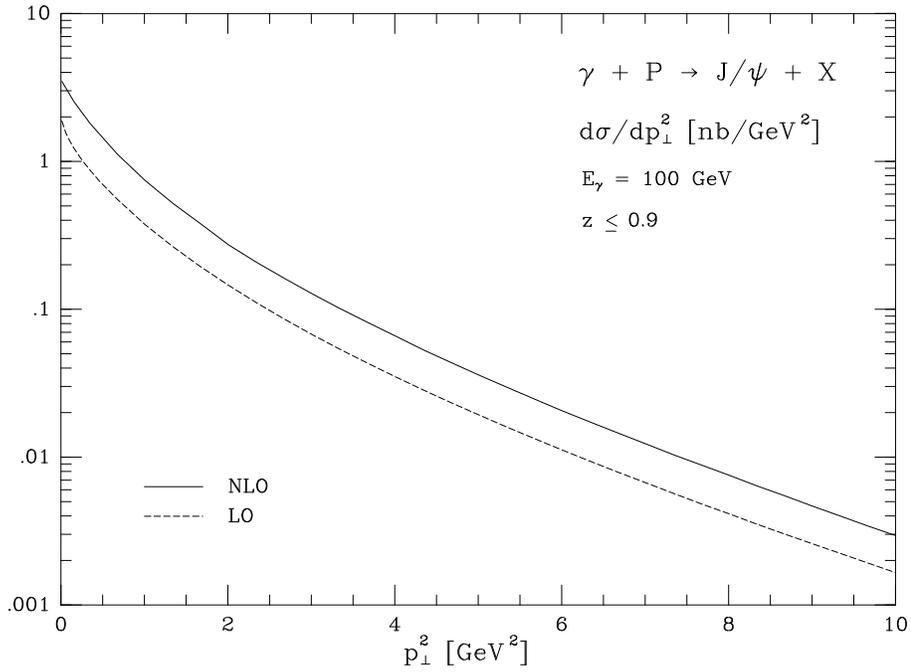,%
height=9cm,%
width=12.5cm,%
bbllx=1.0cm,%
bblly=1.9cm,%
bburx=19.4cm,%
bbury=26.7cm,%
rheight=8.2cm,%
rwidth=15cm,%
angle=-90}

\vspace*{0.5cm}

\caption[ggdiag]{\label{JPPT1}
                 Transverse momentum distribution
                 $\md\sigma/\md{}p_\perp^2$
                 at an initial photon energy of $E_\gamma = 100$~GeV
                 integrated in the inelastic region $z\le 0.9$.}

\end{figure}

In Figs.\ref{JPZ1} and \ref{JPPT1} the $J/\psi$ energy spectrum
d$\sigma/\mbox{d}z$ and the $J/\psi$ transverse momentum distribution
d$\sigma/\mbox{d}p_\perp^2$ are shown at an initial photon energy of
$E_\gamma = 100$~GeV. The GRV parametrizations of the parton densities
\cite{GRV94} have been adopted. They are particularly suited to
%----------------------------------------------------------------------------
%------------------------------------------------------------------------
%----------------------------------------------------------------------------
%----------------------------------------------------------------------------
characterize the magnitude of the radiative corrections since
they allow one to compare the results for the Born cross section
folded with leading-order parton densities, with the cross sections
consistently evaluated for parton cross sections and parton densities
in next-to-leading order. As the average momentum fraction of the
partons $<\!x\!> \sim 0.1$ is moderate, the curves are not sensitive
to the parametrization in the small-$x$ region. The renormalization
scale has been identified with the factorization scale and set to
$\mur = Q = M_{J/\psi}$. For $\alps$ the two-loop formula is used
with $\nlf$ active flavours and
$\Lambda^{(5)}_{\overline{\mbox{\tiny MS}}} = 215$~MeV,
corresponding to the average fit value in Ref.\cite{PDG94}.  Since the
cross section depends strongly on the QCD coupling, we adopt this
measured value, thus allowing for a slight inconsistency to the extend
that the GRV fits are based on a marginally lower value of $\alps$
(cf.\ Ref.\cite{AV95}). In next-to-leading order, the wave-function at the
origin is related to the leptonic $J/\psi$ width by
\beq\label{EQ_DECAYCORR}
\Gamma_{ee} = \left(1 - \frac{16}{3}\frac{\alps}{\pi}\right)
\frac{16\pi\alpha^2e_c^2}{M_{J/\psi}^{2}}\, |\varphi(0)|^2
\eeq
with only transverse gluon corrections taken into account explicitly
\cite{BGR81}. We use $\Gamma_{ee}=5.26$~keV, $M_{J/\psi}=3.097$~GeV
\cite{PDG94} and $m_c=M_{J/\psi}/2$.

{}From Fig.\ref{JPZ1} one can infer that the perturbative QCD analysis
is not under proper control in the limit $z\to 1$, as anticipated for
this singular boundary region. If we restrict the analysis to the
inelastic domain $z\;\simlt\;0.9$ the perturbative expansion is
well-behaved however and the next-to-leading order corrections do not
strongly affect the shape of the distributions.  The $K$-factor,
$K\equiv\sigma_{\mbox{\scriptsize{}NLO}}/\sigma_{\mbox{\scriptsize{}LO}}$,
is nearly independent of $z$ and $p_\perp$ in the inelastic region
$z\;\simlt\; 0.9$. Its magnitude turns out to be $K\sim 2.0$ with one
part $\sim 1.73$ due to the QCD radiative corrections of the leptonic
$J/\psi$ width \cite{BGR81} and a second part $\sim 1.2$ due to the
dynamical QCD corrections. The slope of the transverse momentum
distribution, $\md\sigma/\md{}p_\perp^2 \propto \exp(-b\,p_\perp^2)$,
is predicted to be $b\sim 0.6$~GeV$^{-2}$, in good agreement with the
experimental value $b=0.62\pm 0.2$~GeV$^{-2}$ \cite{NA14}.

\begin{figure}[hbtp]

\hspace*{1.4cm}
\epsfig{%
file=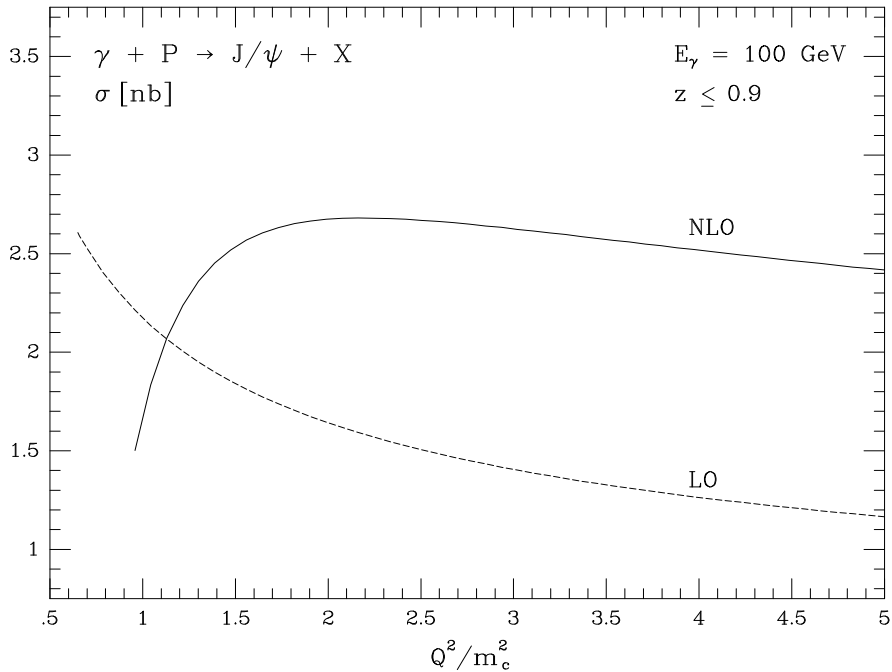,%
height=9.0cm,%
width=12.5cm,%
bbllx=1.0cm,%
bblly=1.9cm,%
bburx=19.4cm,%
bbury=26.7cm,%
rheight=8.2cm,%
rwidth=15cm,%
angle=-90}

\vspace*{0.5cm}

\caption[xx]{
            \label{F_QSQDEP}
            Dependence of the total cross section
            $\gamma + P \to  J/\psi + X$ on the
            renormalization/factorization scale $Q$
            at an initial photon energy of $E_\gamma = 100$~GeV.}

\end{figure}

The scale dependence of the theoretical prediction is reduced
considerably when higher-order QCD corrections are included.
This is demonstrated in Fig.\ref{F_QSQDEP} where we compare the scale
dependence of the leading and next-to-leading order total cross
%---------------------------------------------------------------------------
%-----------------------------------------------------------------
section in the inelastic region $z \le 0.9$. For the sake of
simplicity, the renormalization scale has been identified with the
factorization scale, $\mur = Q$.  While the ratio of the cross
sections in leading order for $Q^2=m_c^2 : (2\,m_c^2) : M_{J/\psi}^2$
is given by $1.7 : 1.3 : 1$, it is much closer to unity, $0.7 : 1.1 :
1$, in the next-to-leading order calculation.  The cross section runs
through a maximum \cite{PMS81} near $Q^2\approx 2{\,}m_c^2$ with broad
width, the origin of the stable behaviour in $Q$.

The next-to-leading order predictions can be confronted with
photoproduction data of the fixed target experiments. For a meaningful
comparison, the theoretical uncertainties due to variation of the
charm quark mass and the strong coupling have to be taken into account
properly. In the static approximation the choice $m_c = M_{J/\psi}/2$
is required for a consistent description of the bound state formation.
A smaller mass value, however, is favoured to describe the charm quark
creation in the hard scattering process \cite{FMNR94}. The value of
the heavy quark mass in the short distance amplitude is the main
parameter controlling the normalization of the cross section. It is
therefore appropriate to adopt charm masses below $m_c =
M_{J/\psi}/2$, thus allowing for a slight correction in the bound
state formation.  Leading-order analyses of $J/\psi$ production that
go beyond the static limit and incorporate a non-vanishing binding
energy, find an effective charm mass value of $m_c = 1.43$~GeV
\cite{JKGW93}, in fairly good agreement with potential model
calculations \cite{BT81}. In order to demonstrate the uncertainty due
to the variation of the charm quark mass, the strong coupling and the
renormalization/factorization scale the results will be shown for (i)
$m_c = M_{J/\psi}/2 \approx 1.55$~GeV with
$\Lambda^{(5)}_{\overline{\mbox{\tiny MS}}} = 215$~MeV and
$Q^2=\mursq=M_{J/\psi}^2$ (as used in the previous figures) and (ii)
$m_c = 1.4$~GeV with $\Lambda^{(5)}_{\overline{\mbox{\tiny MS}}}
= 300$~MeV (corresponding to the 1~$\sigma$ upper boundary of the
error band in Ref.\cite{PDG94}) and $Q^2=\mursq=2\,m_c^2$.

\begin{figure}[hbtp]

\hspace*{1.4cm}
\epsfig{%
file=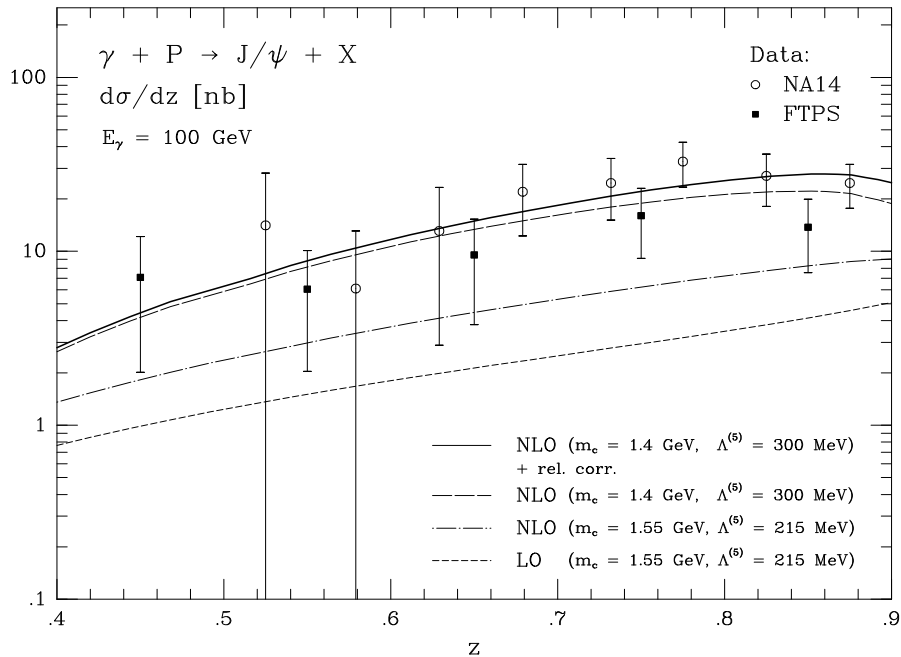,%
height=9cm,%
width=12.5cm,%
bbllx=1.0cm,%
bblly=1.9cm,%
bburx=19.4cm,%
bbury=26.7cm,%
rheight=8.2cm,%
rwidth=15cm,%
angle=-90}

\vspace*{0.25cm}

\caption[xx]{
                 \label{F_ZDISTEXP}
                 Energy spectrum $\md\sigma/\md{}z$,
                 at the initial photon energy $E_\gamma =
                 100$~GeV compared with the photoproduction
                 data \cite{NA14,FTPS}.}

\vspace*{-1.5mm}

\end{figure}

In Fig.\ref{F_ZDISTEXP} we confront the leading and next-to-leading
order calculations with the $J/\psi$ energy spectra measured at photon
energies near $E_\gamma = 100$~GeV \cite{NA14,FTPS}.
%-----------------------------------------------------------------
%-----------------------------------------------------------------
It is clear from Fig.\ref{F_ZDISTEXP} that the variation of the charm
mass and the strong coupling does not strongly affect the shape of the
distribution but only results in some uncertainty concerning the overall
normalization. In a systematic expansion one may finally add the
relativistic corrections as estimated in Ref.\cite{JKGW93}.

\begin{figure}[hbtp]

\hspace*{1.4cm}
\epsfig{%
file=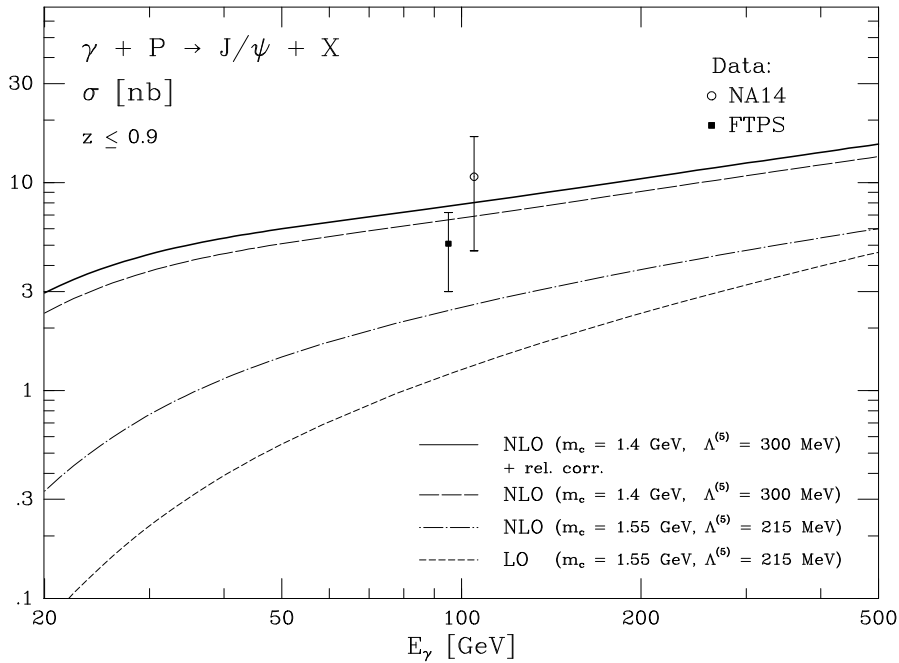,%
height=9cm,%
width=12.5cm,%
bbllx=1.0cm,%
bblly=1.9cm,%
bburx=19.4cm,%
bbury=26.7cm,%
rheight=8.2cm,%
rwidth=15cm,%
angle=-90}

\vspace*{0.5cm}

\caption[ggdiag]{\label{EABHGP1}
                 The total cross section [$z\le 0.9$] as a function
                 of the initial photon energy.}
\end{figure}

The dependence of the total cross section $\gamma + P \to J/\psi + X$
on the photon energy $E_\gamma$ is presented in Fig.\ref{EABHGP1},
again for the two choices of parameters (i) and (ii) as defined above,
together with the photoproduction data \cite{NA14,FTPS}.  From the
curves shown in Fig.\ref{EABHGP1} we deduce that the QCD corrections
are large at moderate photon energies, but decrease with increasing
energies, a consequence of the negative dip in the $c^{(1)}$ scaling
function of Fig.\ref{F_SCALE}.
%---------------------------------------------------------------------
%-----------------------------------------------------------------

Two conclusions can be drawn from the comparison of the
next-to-leading order results with the experimental data. (i) The
$J/\psi$ energy dependence d$\sigma/\mbox{d}z$ and the slope of the
transverse momentum distribution d$\sigma/\mbox{d}p_\perp^2$ are
adequately accounted for by the theoretical prediction in the
inelastic region. (ii) The absolute normalization of the cross section
is somewhat less certain. However, taking into account the theoretical
uncertainty due to variation of the charm quark mass and the strong
coupling and allowing for higher-twist uncertainties of order
$(\Lambda/m_c)^k \;\simlt\; 30\%$ for $k \ge 1$, we conclude that the
normalization too appears to be under semi-quantitative control.

\subsection{Inelastic $J/\psi$ photoproduction at HERA}
The production of $J/\psi$ particles in high energy $ep$ collisions at
\mbox{HERA} is dominated by photoproduction events where the electron
is scattered by a small angle producing photons of almost zero
virtuality. The measurements at \mbox{HERA} provide information on the
dynamics of inelastic $J/\psi$ photoproduction in a kinematical region
very different from that available at fixed target experiments.  The
$\gamma p$ centre of mass energies accessible at \mbox{HERA} are in
the range $30~\mbox{GeV}\;\simlt\;\sqrt{s\hphantom{tk}}\!\!\!\!\!
_{\gamma p}\;\simlt\; 200~\mbox{GeV}$, corresponding to initial photon
energies in a fixed-target experiment of
$450~\mbox{GeV}\;\simlt\;E_\gamma\;\simlt\; 20,000~\mbox{GeV}$.

\begin{figure}[htbp]

\hspace*{1.4cm}
\epsfig{%
file=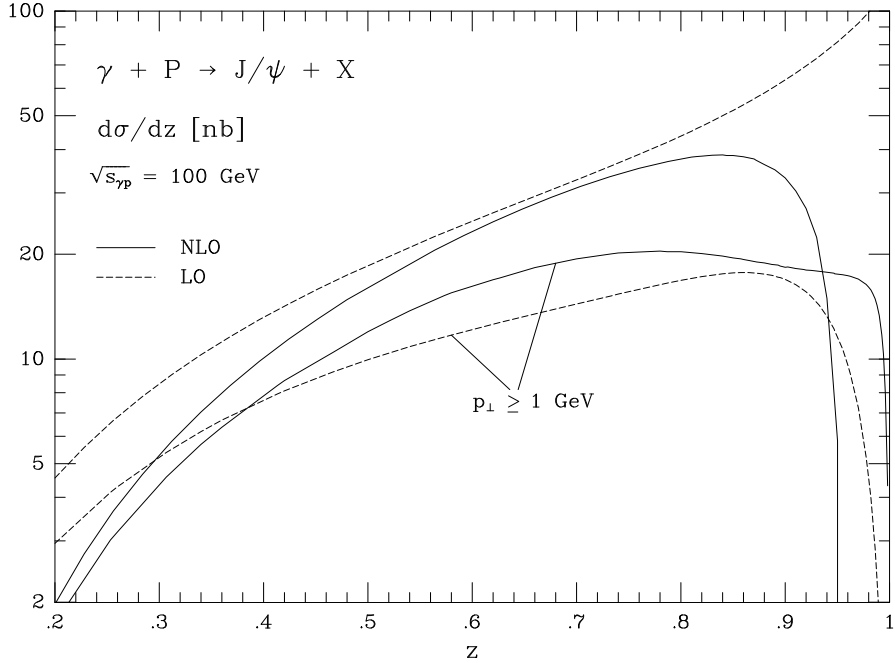,%
height=9cm,%
width=12.5cm,%
bbllx=1.0cm,%
bblly=1.9cm,%
bburx=19.4cm,%
bbury=26.7cm,%
rheight=8.2cm,%
rwidth=15cm,%
angle=-90}

\vspace*{0.5cm}

\caption[ggdiag]{\label{JPZ2}
                 Energy distribution $\md\sigma/\md{}z$
                at the photon-proton centre of mass energy
           $\sqrt{s\hphantom{tk}}\!\!\!\!\! _{\gamma p}\,\, = 100$~GeV
                 integrated in the full $p_\perp$ range and in the
                 restricted range $p_\perp \ge 1$~GeV.}

\end{figure}

\begin{figure}[htbp]

\hspace*{1.4cm}
\epsfig{%
file=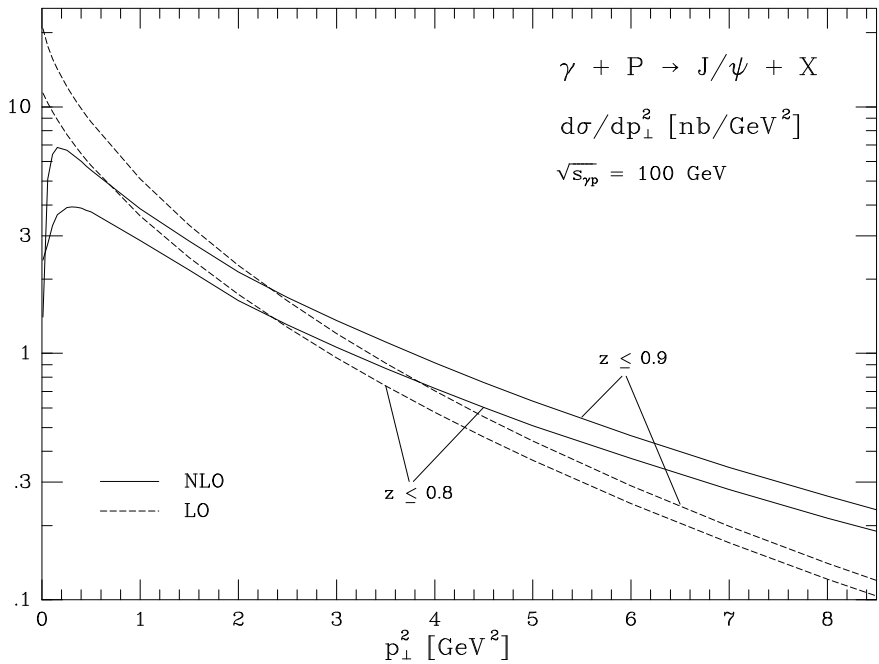,%
height=9cm,%
width=12.5cm,%
bbllx=1.0cm,%
bblly=1.9cm,%
bburx=19.4cm,%
bbury=26.7cm,%
rheight=8.2cm,%
rwidth=15cm,%
angle=-90}

\vspace*{0.5cm}

\caption[ggdiag]{\label{JPPT2}
                 Transverse momentum distribution
                 $\md\sigma/\md{}p_\perp^2$
                at the photon-proton centre of mass energy
           $\sqrt{s\hphantom{tk}}\!\!\!\!\! _{\gamma p}\,\, = 100$~GeV
                 integrated in the region $z\le 0.9$ and $z\le 0.8$.}

\end{figure}

To begin with, we discuss the $J/\psi$ energy spectrum
d$\sigma/\mbox{d}z$ and the $J/\psi$ transverse momentum distribution
d$\sigma/\mbox{d}p_\perp^2$ at a typical \mbox{HERA} energy of
$\sqrt{s\hphantom{tk}}\!\!\!\!\!  _{\gamma p} = 100~\mbox{GeV}$.  The
parameters have been chosen as in the corresponding figures of
Sec.\ref{SEC51}. From Figs.\ref{JPZ2} and \ref{JPPT2} one can conclude
that the next-to-leading order corrections are dominated by strong
negative contributions in the limit $z\to 1$ and $p_\perp\to 0$. This
behaviour which has already been observed in the low energy region of
the fixed-target experiments is much more pronounced in the high
energy range at HERA.  Even if the analysis is restricted to the
region $z\le 0.8$ we still find that the fixed-order perturbative QCD
calculation is not under proper control for $p_\perp \to 0$.  This can
be inferred from Fig.\ref{JPPT2} where the transverse momentum
%----------------------------------------------------------------------------
%------------------------------------------------------------------------
%----------------------------------------------------------------------------
%----------------------------------------------------------------------------
spectrum is shown integrated in the range $z\le 0.9$ and $z\le 0.8$,
respectively. For $p_\perp\;\simlt\; 0.5$~GeV the results of this
calculation obviously require missing contributions from even higher
orders in the perturbative expansion. No reliable prediction can be
made in the small $p_\perp$ and large $z$ domain without resummation
of large logarithmic corrections caused by multiple gluon emission.
It is therefore appropriate to exclude the region $z\to 1$ and
$p_\perp\to 0$ from the analysis. In the following we will present the
results in two kinematic domains, (I)~$z\le 0.9$, which is the minimal
restriction in order to eliminate elastic/diffractive contributions,
and (II)~$z\le 0.8$ and $p_\perp \ge 1$~GeV, which was found to be the
region where fixed-order perturbation theory allows for a reliable
prediction in the \mbox{HERA} energy range. The next-to-leading order
results for the $J/\psi$ energy spectrum are shown in Fig.\ref{JPZ2}
integrated in the full $p_\perp$ range and in the restricted range
$p_\perp \ge 1$~GeV. No singular behaviour is observed for the latter
curve even in the limit $z\to 1$.

\begin{figure}[htbp]

\hspace*{1.4cm}
\epsfig{%
file=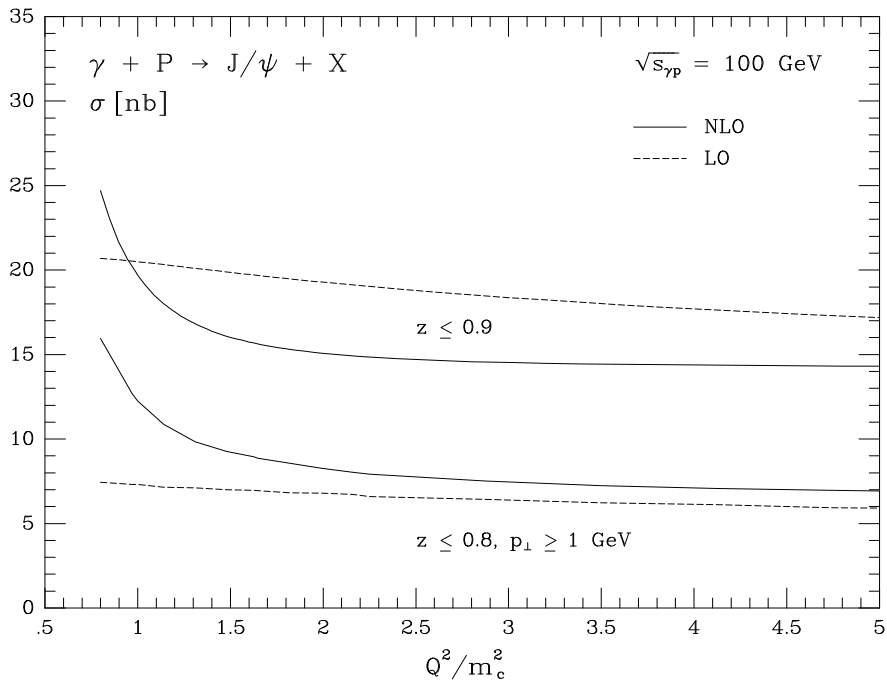,%
height=9.0cm,%
width=12.5cm,%
bbllx=1.0cm,%
bblly=1.9cm,%
bburx=19.4cm,%
bbury=26.7cm,%
rheight=8.2cm,%
rwidth=15cm,%
angle=-90}

\vspace*{0.5cm}

\caption[xx]{
            \label{F_QSQDEP2}
            Dependence of the total cross section
            $\gamma + P \to  J/\psi + X$ on the
            renormalization/factorization scale $Q$
            at the photon-proton centre of mass energy
           $\sqrt{s\hphantom{tk}}\!\!\!\!\! _{\gamma p}\,\, = 100$~GeV.
            The results are shown in two kinematic domains:
            (I) $z \le 0.9$; (II) $z\le 0.8$ and $p_\perp \ge 1$~GeV.}

\end{figure}

In Fig.\ref{F_QSQDEP2} we compare the scale dependence of the
leading order and next-to-leading order total cross sections at the
invariant energy $\sqrt{s\hphantom{tk}}\!\!\!\!\!{}_{\gamma p} =
100~\mbox{GeV}$.
%---------------------------------------------------------------------------
%-----------------------------------------------------------------
The renormalization scale has been identified with the factorization
scale, $\mur=Q$. The results are shown in the two kinematic domains
(I)~$z\le 0.9$ and (II)~$z\le 0.8$ with $p_\perp \ge 1$~GeV, as
discussed above.  From Fig.\ref{F_QSQDEP2} one can infer that the
next-to-leading order result is insensitive to scale variations in an
appreciable range near the reference scale
$\mursq=Q^{2}=M_{J/\psi}^{2}$. For scales below $\mursq=Q^{2}\sim
M_{J/\psi}^{2}/2$ no stable prediction is possible. In contrast to the
low energy region, the cross section in the \mbox{HERA} energy range
does not exhibit a point of minimal scale sensitivity. In the BLM
scheme \cite{BLM83} we find a value of $\mursq\sim M_{J/\psi}^{2}/2$.
This scale is significantly larger than the corresponding BLM value
for $J/\psi$ decays. The typical kinematical energy scale is not set
any more by the small gluon energy in the $J/\psi$ decay but rather by
the typical initial-state parton energies.

\begin{figure}[htbp]

\hspace*{1.4cm}
\epsfig{%
file=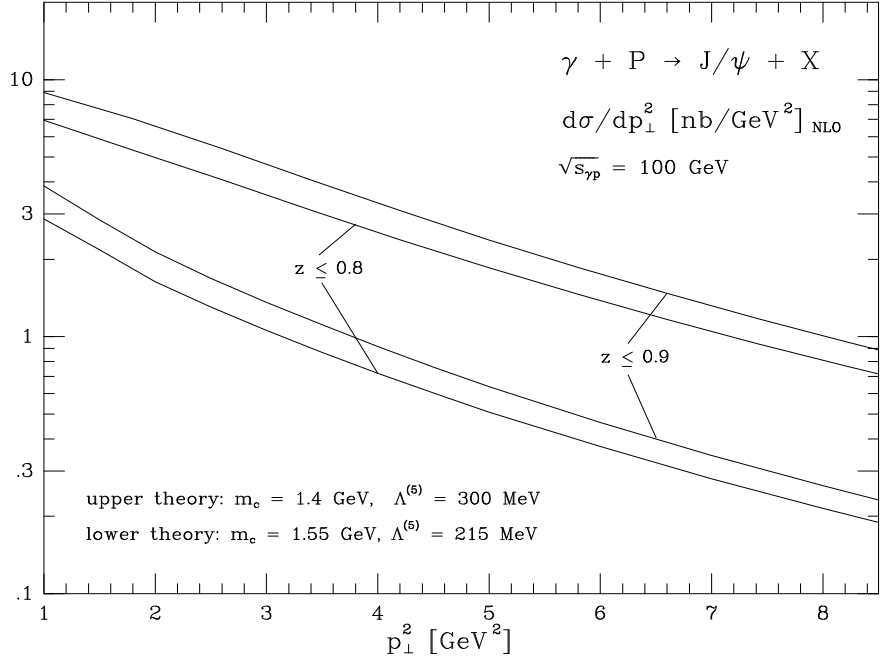,%
height=9cm,%
width=12.5cm,%
bbllx=1.0cm,%
bblly=1.9cm,%
bburx=19.4cm,%
bbury=26.7cm,%
rheight=8.2cm,%
rwidth=15cm,%
angle=-90}

\vspace*{0.5cm}

\caption[ggdiag]{\label{JPPT3}
                 Transverse momentum distribution
                 $\md\sigma/\md{}p_\perp^2$
                at the photon-proton centre of mass energy
           $\sqrt{s\hphantom{tk}}\!\!\!\!\! _{\gamma p}\,\, = 100$~GeV
                 integrated in the region $z\le 0.9$ and $z\le 0.8$.}

\end{figure}

\begin{figure}[htbp]

\hspace*{1.4cm}
\epsfig{%
file=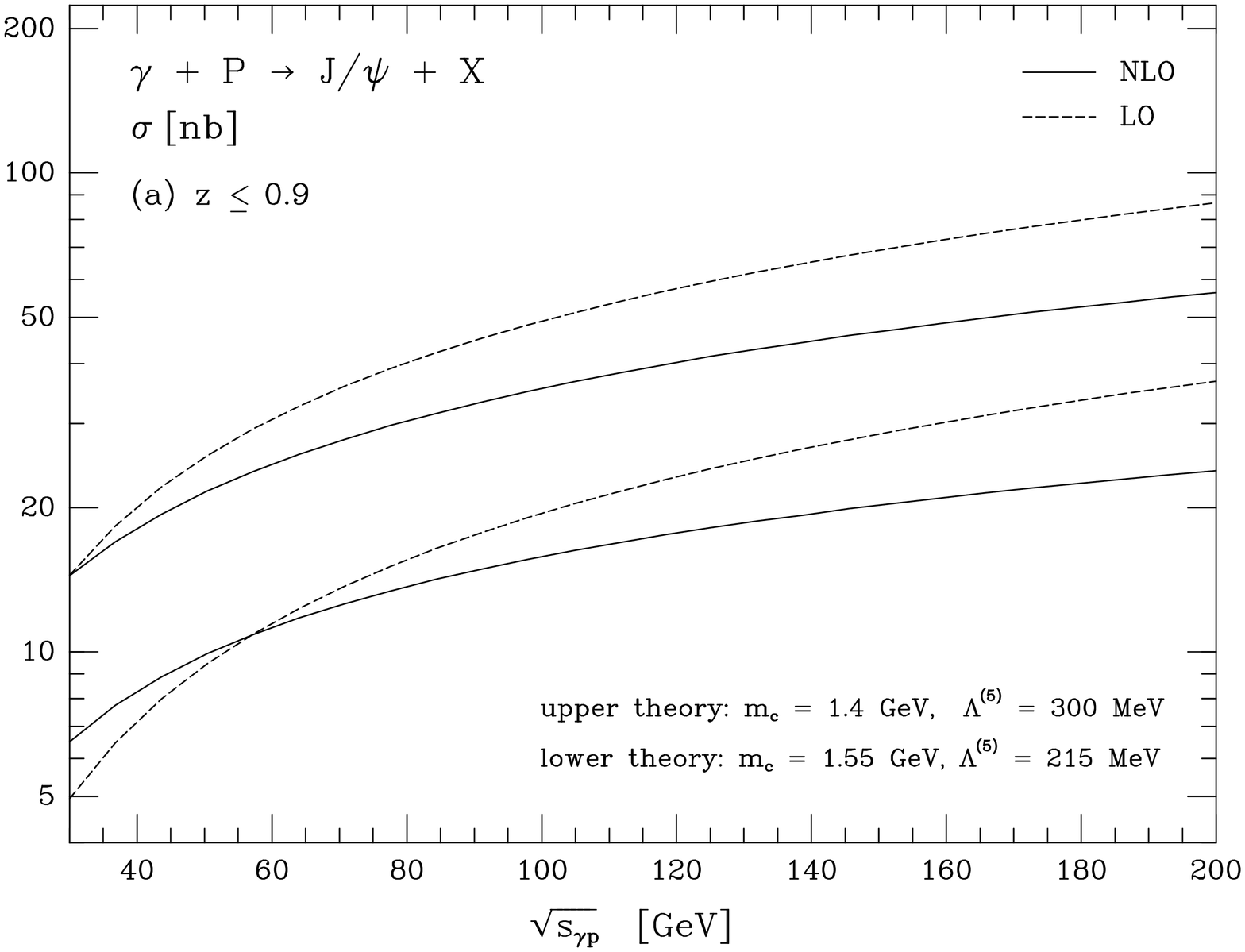,%
height=9.0cm,%
width=12.5cm,%
bbllx=1.0cm,%
bblly=1.9cm,%
bburx=19.4cm,%
bbury=26.7cm,%
rheight=8.2cm,%
rwidth=15cm,%
angle=-90}

\vspace*{2cm}

\hspace*{1.4cm}
\epsfig{%
file=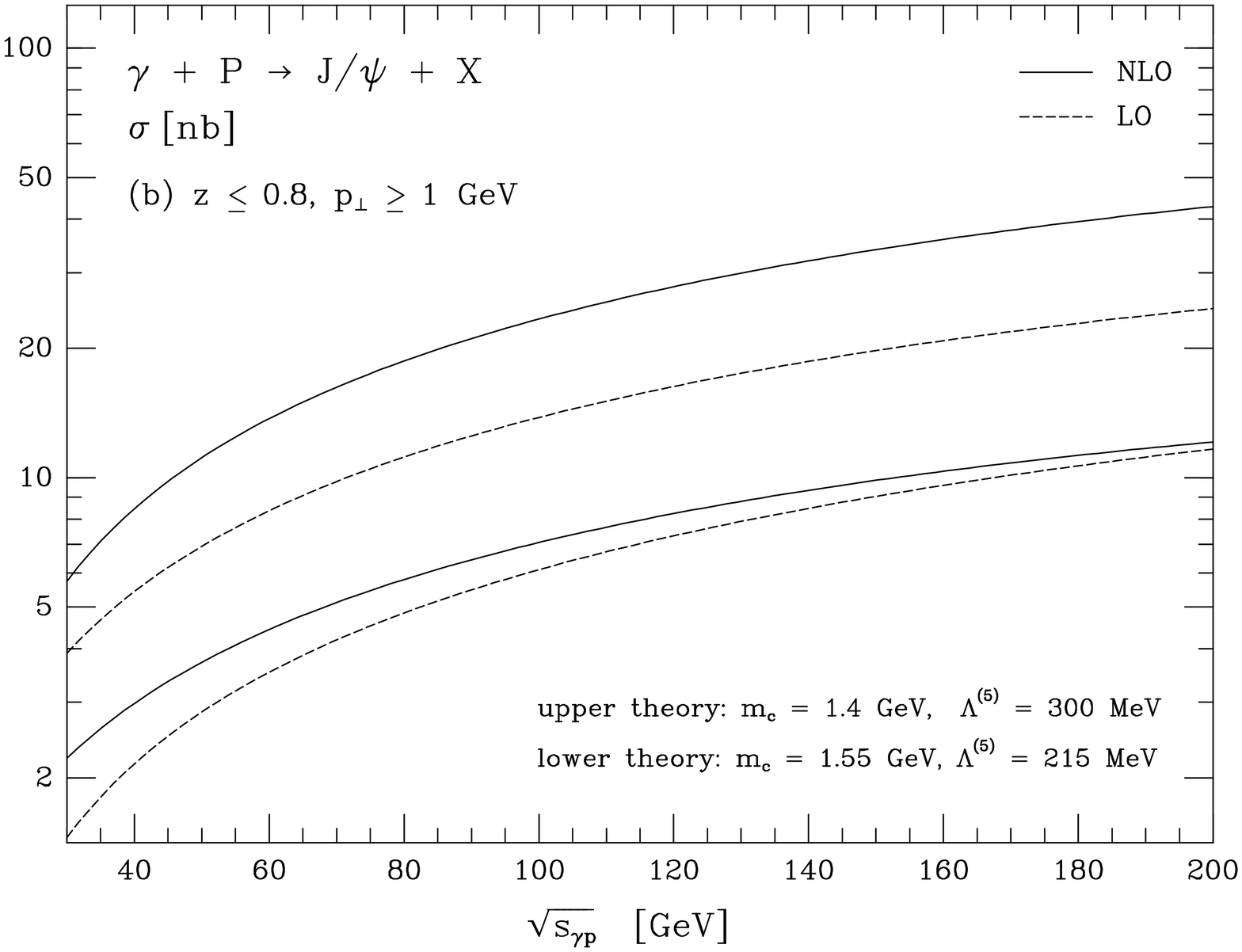,%
height=9.0cm,%
width=12.5cm,%
bbllx=1.0cm,%
bblly=1.9cm,%
bburx=19.4cm,%
bbury=26.7cm,%
rheight=8.2cm,%
rwidth=15cm,%
angle=-90}

\vspace*{0.6cm}

\caption[xx]{
    \label{F_TOTHERA1}
    The total cross section for inelastic $J/\psi$ photoproduction
    $\gamma+P\to J/\psi+X$ as a function of the photon-proton centre
    of mass energy in the HERA energy range. The results are shown
    in two kinematic domains: (a) $z \le 0.9$; (b) $z \le 0.8$ and
    $p_\perp \ge 1$~GeV.}

\end{figure}

We will now present our final predictions for differential
distributions and total cross sections for inelastic $J/\psi$
photoproduction at \mbox{HERA}. In a systematic expansion we have
added the relativistic corrections \cite{JKGW93} which enhance the
large $z$ and small $p_\perp$ region and thereby increase the total
cross section integrated in the range $z\le 0.9$ by $\approx$~10\%.
The inclusion of relativistic corrections does not change the results
obtained in the more restricted domain $z\le 0.8$ and $p_\perp \ge
1$~GeV.  In order to demonstrate the theoretical uncertainty the
results are shown for (i) $m_c = M_{J/\psi}/2 \approx 1.55$~GeV with
$\Lambda^{(5)}_{\overline{\mbox{\tiny MS}}} = 215$~MeV and
$Q^2=\mursq=M_{J/\psi}^2$ (as used in the previous figures) and (ii)
$m_c = 1.4$~GeV with $\Lambda^{(5)}_{\overline{\mbox{\tiny MS}}}
= 300$~MeV and $Q^2=\mursq=2\,m_c^2$, as discussed in Sec.\ref{SEC51}.
In Fig.\ref{JPPT3} we plot the transverse momentum distribution at the
photon-proton centre of mass energy $\sqrt{s\hphantom{tk}}\!\!\!\!\!
_{\gamma p}\,\, = 100$~GeV integrated in the region $z\le 0.9$ and
%----------------------------------------------------------------------------
%----------------------------------------------------------------------------
$z\le 0.8$. As could already be inferred from Fig.\ref{JPPT2}, the
inclusion of the next-to-leading order corrections increases the cross
section in the range $p_\perp\;\simgt\;1$~GeV and results in a
hardening of the distribution. For the slope,
$\md\sigma/\md{}p_\perp^2 \propto \exp(-b\,p_\perp^2)$, we predict
$b\sim 0.3$~GeV$^{-2}$.  The $J/\psi$ energy distribution
%----------------------------------------------------------------------------
\begin{figure}[htbp]

\hspace*{1.4cm}
\epsfig{%
file=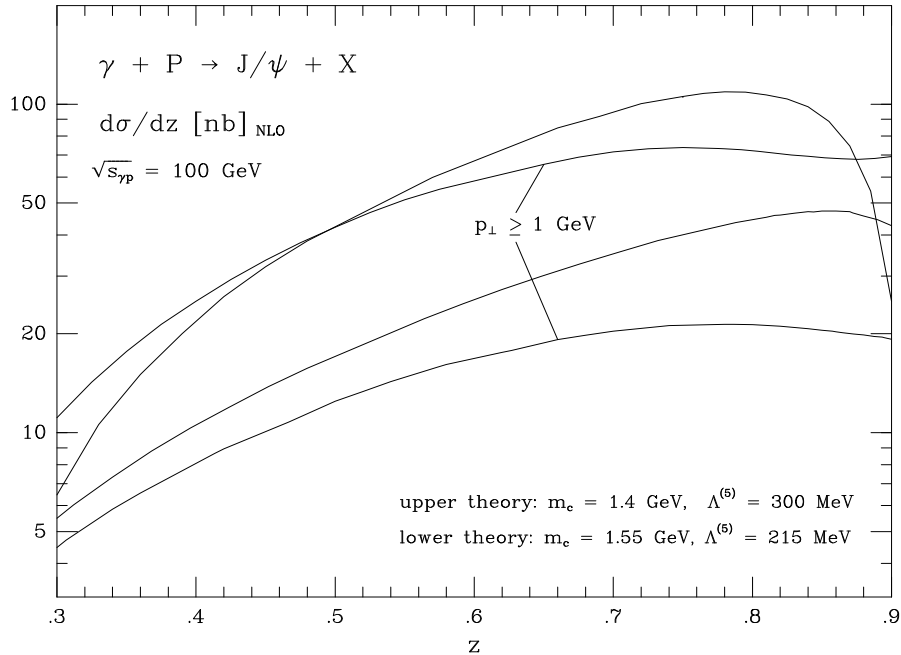,%
height=9cm,%
width=12.5cm,%
bbllx=1.0cm,%
bblly=1.9cm,%
bburx=19.4cm,%
bbury=26.7cm,%
rheight=8.2cm,%
rwidth=15cm,%
angle=-90}

\vspace*{0.5cm}

\caption[ggdiag]{\label{JPZ3}
                 Energy distribution $\md\sigma/\md{}z$
                at the photon-proton centre of mass energy
           $\sqrt{s\hphantom{tk}}\!\!\!\!\! _{\gamma p}\,\, = 100$~GeV
                 integrated in the full $p_\perp$ range and in the
                 restricted range $p_\perp \ge 1$~GeV.}

\end{figure}
%------------------------------------------------------------------------
is shown in Fig.\ref{JPZ3} integrated in the full $p_\perp$ range and
in the restricted range $p_\perp \ge 1$~GeV, again for two choices of
the charm quark mass and the strong coupling.  In
Figs.\ref{F_TOTHERA1}(a) and (b) we plot the total cross section as a
function of the photon-proton centre of mass energy in the \mbox{HERA}
range. The results are shown in the two kinematic domains (I)~$z\le
0.9$ and (II)~$z\le 0.8$ with $p_\perp \ge 1$~GeV. For domain (I),
Fig.\ref{F_TOTHERA1}(a), the $K$-factor is $\sim 0.75$, a consequence
of the strong negative contribution present in the region $p_\perp \to
0$.  In the more restricted kinematic domain (II),
Fig.\ref{F_TOTHERA1}(b), the next-to-leading order corrections
significantly increase the cross section. For $m_c = 1.4$~GeV with
$\Lambda^{(5)}_{\overline{\mbox{\tiny MS}}} = 300$~MeV and
$Q^2=\mursq=2\,m_c^2$, we find
$K\equiv\sigma_{\mbox{\scriptsize{}NLO}}/\sigma_{\mbox{\scriptsize{}LO}}
\approx 1.7$.
%---------------------------------------------------------------------------
%---------------------------------------------------------------------------

An estimate of the cross section for inelastic photoproduction of
$\psi'$ particles can be obtained from the results presented here
by replacing the leptonic decay width and multiplying with a phase
space correction factor
\bea\label{PSII}
 \sigma(\gamma P \to \psi'\; X) & \approx & \Gamma_{ee}^{\psi'} /
\Gamma_{ee}^{J/\psi}\; (M_{J/\psi}/M_{\psi'})^3 \times \sigma(\gamma
P\to J/\psi\;X)\nonumber\\
& \approx & 1/4 \times \sigma(\gamma P\to J/\psi\;X)
\quad .
\eea
The estimate (\ref{PSII}) should be considered as a lower bound since
it is based on a purely static approach. In the derivation of the
phase space suppression factor $(M_{J/\psi}/M_{\psi'})^3$ it is
assumed that the effective charm masses in the short distance
amplitudes scale like the corresponding $\psi'$ and $J/\psi$ masses.
This difference might be reduced by including relativistic
corrections which are expected to be significantly larger for the
$\psi'$ state than for the $J/\psi$ \cite{BT81}. A measurement
of the inelastic $\psi'$ photoproduction cross section at \mbox{HERA}
would clearly help to understand the impact of relativistic
corrections on charmonium production.

The production of $\Upsilon$ bottomonium bound states is suppressed,
compared with $J/\psi$ states, by a factor of about 300 at
\mbox{HERA}, a consequence of the smaller bottom electric charge and
the phase space reduction by the large $b$ mass.

\begin{figure}[htbp]

\hspace*{1.4cm}
\epsfig{%
file=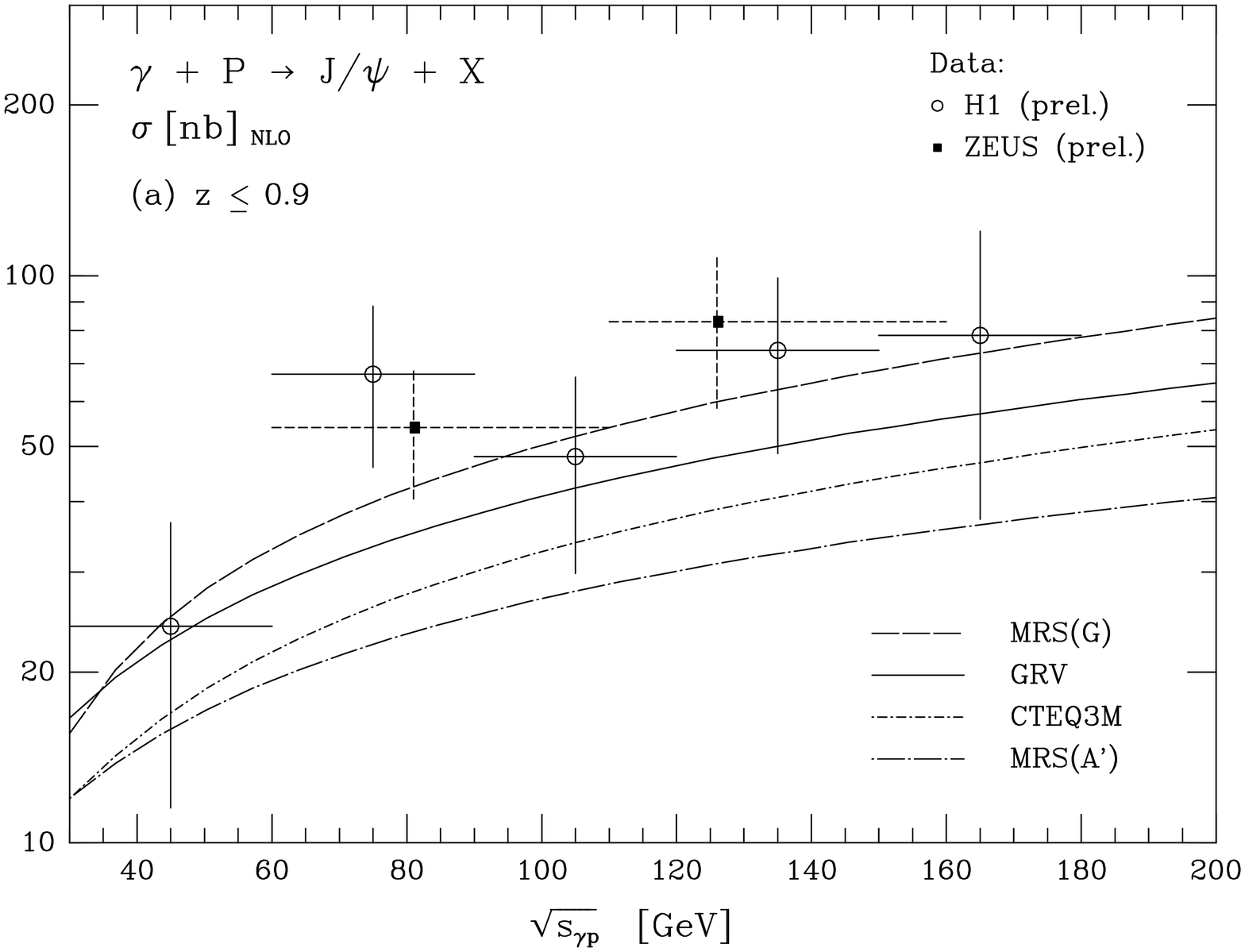,%
height=9.0cm,%
width=12.5cm,%
bbllx=1.0cm,%
bblly=1.9cm,%
bburx=19.4cm,%
bbury=26.7cm,%
rheight=8.2cm,%
rwidth=15cm,%
angle=-90}

\vspace*{2cm}

\hspace*{1.4cm}
\epsfig{%
file=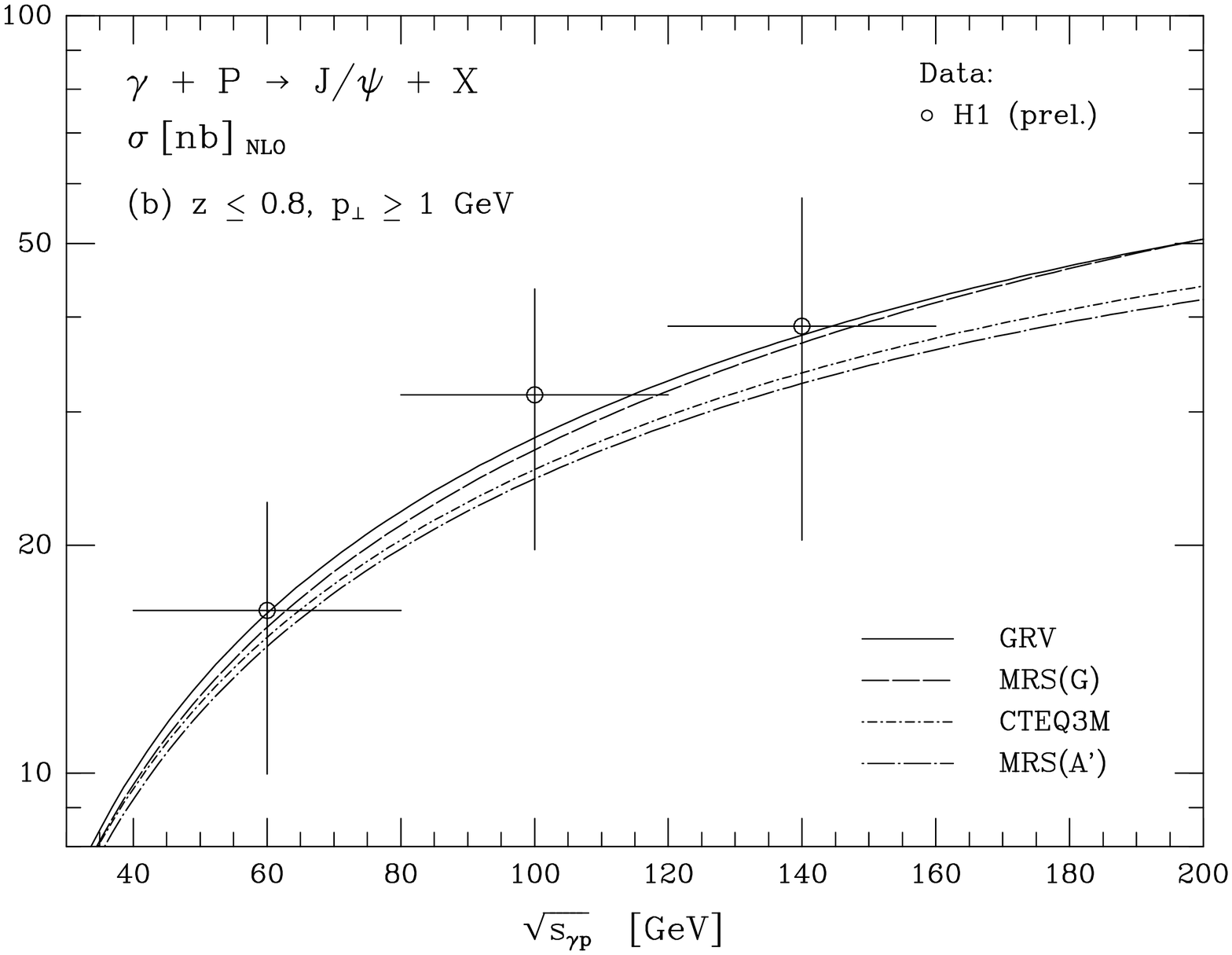,%
height=9.0cm,%
width=12.5cm,%
bbllx=1.0cm,%
bblly=1.9cm,%
bburx=19.4cm,%
bbury=26.7cm,%
rheight=8.2cm,%
rwidth=15cm,%
angle=-90}

\vspace*{0.6cm}

\caption[xx]{
    \label{F_TOTHERA2}
    The total cross section as a function of the photon-proton
    centre of mass energy for different parametrizations of the
    gluon distribution of the proton compared with preliminary
    data from \mbox{H1} \cite{H1PSI} and \mbox{ZEUS} \cite{ZEUSPSI}.
    The results are shown in two kinematic domains:
    (a) $z \le 0.9$; (b) $z \le 0.8$ and $p_\perp \ge 1$~GeV.}

\end{figure}

Since the momentum fraction of the partons at \mbox{HERA} energies is
small, the cross sections presented above are sensitive to the
parametrization of the gluon distribution in the small-$x$ region $<\!
x\! > \sim 0.003$. In Figs.\ref{F_TOTHERA2}(a) and (b) we compare the
next-to-leading order predictions for different parametrizations of
the gluon distribution in the proton with first results measured at
\mbox{HERA} \cite{H1PSI,ZEUSPSI}. We have used $m_c = 1.4$~GeV with
%---------------------------------------------------------------------------
%-------------------------------------------------------------------------
$\Lambda^{(5)}_{\overline{\mbox{\tiny MS}}} = 300$~MeV and
$Q^2=\mursq=2\,m_c^2$, as favoured by the low energy data. The data
samples \cite{H1PSI,ZEUSPSI} contain background events from production
of $\psi'$ states with subsequent $J/\psi$ decay \cite{BRUG}. This
contribution has been included in the theoretical prediction,
Figs.\ref{F_TOTHERA2}(a) and (b), in order to allow for a meaningful
comparison. According to (\ref{PSII}) and the measured branching ratio
$\mbox{BR}(\psi'\to J/\psi\,X)=57$~\% \cite{PDG94} the cascade
production from $\psi'$ states is conservatively expected to increase
the cross section by $\approx$~15\%.  For the parametrizations of the
parton densities in the proton we have chosen four sets that are
compatible with the recent \mbox{HERA} measurements of the proton
structure functions \cite{HERAF2}: the GRV parametrization
\cite{GRV94}, which has been adopted in all previous figures, the sets
MRS(G), MRS(A') \cite{MRS95} and the CTEQ3 parametrization
\cite{CTEQ}. The difference in the small
$x$ behaviour of the gluon densities, $xg \sim x^{-\lambda}$ where
$\lambda_{\mbox{\scriptsize GRV}} \approx 0.3-0.4$,
$\lambda_{\mbox{\scriptsize MRS(G)}} \approx 0.4$,
$\lambda_{\mbox{\scriptsize MRS(A')}} \approx 0.2$ and
$\lambda_{\mbox{\scriptsize CTEQ3M}} \approx 0.3$, results in
different normalizations and, to a smaller extend, in different shapes
of the cross section as a function of the photon-proton centre of mass
energy. The shape of the differential distributions in $z$ and
$p_\perp$ is not very sensitive to changes in the proton structure
function. The theoretical result for the total cross section
integrated in the kinematic domain $z\le 0.9$ on average
underestimates the experimentally observed production rate,
Fig.\ref{F_TOTHERA2}(a).  This could be anticipated since the
perturbative analysis is not under proper control in the limit
$p_\perp \to 0$ and the next-to-leading order corrections are
dominated by strong negative contributions for
$p_\perp\;\simlt\;0.5$~GeV, as shown in Fig.\ref{JPPT2}.
Following the arguments above, it is more adequate to compare theory
and experiment in the kinematic domain $z\le 0.8$ and $p_\perp \ge
1$~GeV, where fixed-order perturbation theory allows for a reliable
prediction in the \mbox{HERA} energy range. This comparison is shown
in Fig.\ref{F_TOTHERA2}(b). It can be inferred from the plot that the
next-to-leading order result not only accounts for the energy
dependence of the cross section but also for the overall
normalization. The sensitivity of the prediction to the gluon
distribution in the proton is, however, not very distinctive in the
more restricted domain $z\le 0.8$ and $p_\perp \ge 1$~GeV.  In
particular the MRS(G) and GRV parton densities lead to almost
identical results over the whole kinematical range accessible at
\mbox{HERA}.  A detailed analysis reveals that the size of the QCD
corrections increases when adopting parton densities with flatter
gluons.  The sensitivity to different gluon distributions is thus
reduced in next-to-leading order as compared to the leading-order
result, in particular when choosing a small charm mass and a large
value for the strong coupling. Parametrizations with extremely flat
gluons like MRS(D0') \cite{MRSD0} are clearly disfavoured by the
recent \mbox{HERA} measurements of the proton structure function
\cite{HERAF2} and do not allow for a reliable prediction in the high
energy region.  For the parameters adopted in Fig.\ref{F_TOTHERA2}(b),
the MRS(D0') distribution leads to next-to-leading order results not
very different from those obtained with the MRS(A') parametrization.
The corresponding $K$-factors are however uncomfortably large, $K \sim
4$, casting doubts on the reliability of the perturbative expansion as
obtained by using flat gluon distributions. If parton distributions
with steep gluon densities are adopted, the next-to-leading order
cross section is well-behaved and gives an adequate description of the
experimental data, as demonstrated in Fig.\ref{F_TOTHERA2}.

\section{Conclusion}

We have presented a complete calculation of the higher-order
perturbative QCD corrections to inelastic photoproduction of $J/\psi$
particles. In the energy range of the fixed target experiments,
$E_\gamma \sim 100$~GeV, including the next-to-leading order
corrections reduces the scale dependence of the theoretical prediction
and increases the cross section by about a factor of two. A comparison
with photoproduction data of fixed-target experiments reveals that the
$J/\psi$ energy dependence and the slope of the transverse momentum
distribution are adequately accounted for by the theoretical
prediction in the inelastic region $z\;\simlt\;0.9$. Taking into
account the uncertainty due to variation of the charm quark mass and
the strong coupling and allowing for higher-twist effects of order
$(\Lambda/m_c) \;\simlt\; 30\%$, we conclude that the normalization
too appears to be under semi-quantitative control. In the high energy
range at \mbox{HERA}, a detailed analysis of the spectra has shown
that the perturbative calculation is not well-behaved in the limit
$p_\perp \to 0$. No reliable prediction can be made in this singular
boundary region without resummation of large logarithmic corrections
caused by multiple gluon emission. First experimental results from
\mbox{HERA} indeed indicate that the production rate, obtained in the
full $p_\perp$ range, is on average underestimated by the theoretical
predictions. If the small $p_\perp$ region is excluded from the
analysis, the next-to-leading order result not only accounts for the
energy dependence of the cross section but also for the overall
normalization. The results seem to favour a gluon density in the
proton rising toward low $x$, consistent with recent measurements of
the proton structure functions. Higher-twist effects must be included
to improve the quality of the theoretical analysis further.

\pagebreak[3]

{\noindent\bf Acknowledgements}\\[1mm]
It is a pleasure to thank J.~Zunft for a fruitful collaboration during
earlier stages of this project and P.M.~Zerwas for valuable advice,
discussions and suggestions. I also thank J.G.~K\"{o}rner for comments
and support. I have benefitted from helpful suggestions concerning
technical aspects of the calculation by W.~Beenakker and M.~Spira and
from discussions with M.~Cacciari, M.~Greco, R.~H\"{o}pker, W.~Kilian,
J.H.~K\"{u}hn, R.~R\"{u}ckl, W.J.~Stirling, W.L.~van~Neerven and
A.~Vogt.  I thank J.~Steegborn for interesting conversations and
comparison of part of the results. Last but not least, I would like to
acknowledge discussions on experimental aspects of $J/\psi$
photoproduction with S.~Bertolin, R.~Brugnera, H.~Jung, B.~Naroska,
S.~Schieck, G.~Schmidt, R.~Sell and L.~Stanco.

\vspace*{1cm}

\appendix

\noindent
{\large \bf APPENDICES}

\section{The scalar integrals}\label{APP1}
Analytical results for the scalar integrals which emerge from the
Passarino-Veltman reduction of the virtual amplitude are listed in
this Appendix. The exchange of longitudinal gluons between the massive
quarks in diagram Fig.\ref{BOXES}b leads to a Coulomb singularity
$\sim \pi^2/v$, which appears in the evaluation of the corresponding
loop integrals.  The singularity has been isolated by introducing a
small quark velocity $v$, as discussed in detail below. First, we will
give the expressions for the Coulomb-finite integrals. The results are
{\em not} analytically continued into the physical region
\beq
s\,,\;s_1+4m^2\ge 4m^2, \phantom{kkkkk}
t\,,\;t_1+4m^2\le 0,  \phantom{kkkkk}
u\,,\;u_1+4m^2\le 0
\quad ,
\eeq
with $s_1$, $t_1$ and $u_1$ as defined in (\ref{INVDEF1}). Therefore,
the expressions for all permutations of the photon and gluon momenta
can be obtained from the formulae listed below by interchanging
$t_1\leftrightarrow u_1$, $s_1\leftrightarrow t_1$, or
$s_1\leftrightarrow u_1$. Since any imaginary part of the integrals
will disappear in the final result, only the real parts are given.
The four-momenta are related by $k_1+k_2=2p+k_3$ and all particles are
taken to be on-mass-shell, $k_1^2 = k_2^2 = k_3^2 = 0$ and $p^2=m^2$.

\noindent
$-$ The scalar one-point function is given by
\beq
A(m^2) = \mu^{4-n}\;\int\frac{\md^nq}{(2\pi)^n}\;
\frac{1}{q^2-m^2} =
iC_{\epsilon}\;m^2\,\left[\,\frac{1}{\epsilon}+1\,\right]
\eeq
where
\beq\label{EQ_CEPS}
C_{\epsilon} = \frac{1}{16\pi^2} \;\; e^{-\epsilon(\gamma_E-\ln 4\pi)}
           \left(\frac{\mu^2}{m^2}\right)^{\epsilon}
\eeq
and $n=4-2\epsilon$.

\noindent
$-$ The scalar two-point function is defined by
\beq
B(p,m_1,m_2) = \mu^{4-n}\;\int\frac{\md^nq}{(2\pi)^n}\;
\frac{1}{[q^2-m_1^2]\,[(q+p)^2-m_2^2]}
\quad
\eeq
The following types of two-point functions appear in the
calculation of the virtual amplitude:
\bea
& & \hspace*{-1.4cm}
B(k_1,0,0)  \:=\:  0 \\[1mm]
& & \hspace*{-1.4cm}
B(k_2-k_3,0,0)  \:=\:
      iC_{\epsilon}\;\left[\,\frac{1}{\epsilon}
       -\ln\left(-\frac{t}{m^2}\right)+2\,\right] \\[1mm]
& & \hspace*{-1.4cm}
B(p,0,m) \:=\: B(2p,m,m) \:=\:
      iC_{\epsilon}\;\left[\,\frac{1}{\epsilon}+2\,\right] \\[1mm]
& & \hspace*{-1.4cm}
B(p-k_1,0,m)  \:=\:  iC_{\epsilon}\;\left[\,
       \frac{1}{\epsilon}+2-\frac{t_1}{t_1+2m^2}
       \ln\left(\frac{-t_1}{\,2m^2\,}\right)\right] \\[1mm]
& & \hspace*{-1.4cm}
B(k_1+k_2,m,m)  \:=\:  iC_{\epsilon}\;\left[\,
       \frac{1}{\epsilon}+2+\beta\ln\left(\frac{1-\beta}{1+\beta}\right)
       \right]
\quad .
\eea
Here and below we have used the shorthand notation
$\beta\equiv\beta(s)=\sqrt{1-4m^2/s}$.

\noindent
$-$ The scalar three-point function is defined by
\bea
\lefteqn{\hspace*{-1cm}
C(p_1,p_2,m_1,m_2,m_3) = }\nonumber\\[1mm]
& &
\mu^{4-n}\;\int\frac{\md^nq}{(2\pi)^n}\;
\frac{1}{[q^2-m_1^2]\,[(q+p_1)^2-m_2^2]\,[(q+p_2)^2-m_3^2]}
\quad .
\eea
Eight different types of three-point functions appear in the virtual
amplitude:
\bea & & \hspace*{-1.175cm}
C(k_2,k_3,0,0,0)  \:=\:  iC_{\epsilon}\;\frac{1}{t}\;
          \left[\,\frac{1}{\epsilon^2}
             -\frac{1}{\epsilon}\ln\left(\frac{-t}{\,m^2\,}\right)
               +\frac{1}{2}\ln^2\left(\frac{-t}{\,m^2\,}\right)
                -\frac{1}{2}\zeta(2)\right] \\[2mm]
& & \hspace*{-1.175cm}
C(-k_3,-k_3-p,0,0,m)  \:=\:  iC_{\epsilon}\;\frac{1}{s_1}
          \left[
            \vphantom{2\Li_2\left(\frac{s_1+2m^2}{2m^2}\right)}
             \,\frac{1}{\epsilon^2}
             -\frac{2}{\epsilon}\ln\left(\frac{-s_1}{2m^2}\right)
            +2\ln^2\left(\frac{-s_1}{2m^2}\right) \right.
                     \nonumber \\[1mm]
& & \hspace*{5.30cm}
        + \left. 2\Li_2\left(\frac{s_1+2m^2}{2m^2}\right)
                 + \frac{1}{2}\zeta(2)\right] \\[2mm]
& & \hspace*{-1.175cm}
C(k_3-k_2,-p,0,0,m)  \:=\:  iC_{\epsilon}\;\frac{2}{t_1}\;\left[
             \frac{1}{2}\ln^2\left(-\frac{t}{2m^2}\right)
      + \Li_2\left(1-\frac{2m^2}{t}\right)\right.
            \nonumber \\[1mm]
 & & \hspace*{4.09cm}
   \left. {}+ \Li_2\left(\frac{t}{4m^2}\right)
    - \Li_2\left(-\frac{2(2m^2-t)}{t}\right)
    + \frac{5}{2}\zeta(2)\right] \\[2mm]
& & \hspace*{-1.175cm}
C(-k_3,p-k_2,0,0,m)  \:=\:  iC_{\epsilon}\;\frac{2}{u_1-t_1}\;
      \left[\,\frac{1}{\epsilon}\ln\left(\frac{t_1}{u_1}\right)
                        \right.
       +\frac12\ln^2\left(\frac{-u_1}{2m^2}\right)
       -\frac12\ln^2\left(\frac{-t_1}{2m^2}\right) \nonumber \\[1mm]
 & & \hspace*{5.99cm}
   \left.
       {} + \Li_2\left(1+\frac{2m^2}{t_1}\right)
        - \Li_2\left(1+\frac{2m^2}{u_1}\right) \right] \\[2mm]
& & \hspace*{-1.175cm}
C(p,p-k_1,0,m,m)  \:=\:  iC_{\epsilon}\;\frac{2}{t_1}\;
        \left[ -\Li_2\left(1+\frac{t_1}{2m^2}\right)+\zeta(2)\right]
        \\[2mm]
& & \hspace*{-1.175cm}
C(p-k_1,-p-k_3,0,m,m)  \:=\:  iC_{\epsilon}\;\frac{2}{s_1-t_1}\;
        \left[\, \Li_2\left(1+\frac{t_1}{2m^2}\right)
         -  \Li_2\left(1+\frac{s_1}{2m^2}\right)\right] \\[2mm]
& & \hspace*{-1.175cm}
\lefteqn{
C(-p,p+k_3,0,m,m)  \:=\:  iC_{\epsilon}\;\frac{2}{s_1}
        \left[
          \vphantom{2\Li_2\left(\frac{s_1+2m^2}{2m^2}\right)}
          \Li_2\left(1-\frac{s}{2m^2}\right) + \frac{1}{2}\zeta(2) \right. }
                         \nonumber \\[1mm]
 & & \hspace*{3.0cm}
  {} - \Li_2\left(\frac{4m^2-2s}{4m^2-s(1+\beta)}\right)
     + \Li_2\left(\frac{4m^2}{4m^2-s(1+\beta)}\right) \nonumber
                           \\[1mm]
 & & \hspace*{3.0cm}
  {} - \Li_2\left(\frac{4m^2-2s}{4m^2-s(1-\beta)}\right)
     + \Li_2\left(\frac{4m^2}{4m^2-s(1-\beta)}\right)
  \left.\vphantom{\Li_2\left(\frac{4m^2-2s}{4m^2-s(1-\beta)}\right)}
   \right] \\[2mm]
& & \hspace*{-1.175cm}
C(k_1,k_1+k_2,m,m,m)  \:=\: -iC_{\epsilon}\;\frac{1}{s}\;
        \left[\,\Li_2\left(\frac{2}{1+\beta}\right)
         + \Li_2\left(\frac{2}{1-\beta}\right) \right]
\quad ,
\eea
where $\zeta(2) = \pi^2/6$.

\noindent
$-$ The scalar four-point function is defined by
\bea
&& \hspace{-1.3cm}D(p_1,p_2,p_3,m_1,m_2,m_3,m_4) = \nonumber\\[1mm]
&& \hspace{-1.0cm}
\mu^{4-n}\;\int\frac{\md^nq}{(2\pi)^n}\;
\frac{1}{[q^2-m_1^2]\,[(q+p_1)^2-m_2^2]\,[(q+p_2)^2-m_3^2]
\,[(q+p_3)^2-m_4^2]}
\quad .
\eea
There are three different types of four-point functions:
\bea & & \hspace*{-2.0cm}
D(k_2,k_3,-p+k_2,0,0,0,m) =
           iC_{\epsilon}\;\frac{2}{t u_1}
           \left\{\,\frac{3}{2}\frac{1}{\epsilon^2}
           - \frac{1}{\epsilon}\,\left[\,
           2\ln\left(\frac{-u_1}{\,2m^2\,}\right)
           -\ln\left(\frac{-t_1}{\,2m^2\,}\right)
%           \vphantom{\Li_2\left(\frac{s_1+u_1}{s_1}
%                     \frac{t_1-s_1}{2(t_1+4m^2)}\right)}
                              \right. \right. \nonumber \\[1mm]
& & \hspace*{3cm}
  {}   \left. +\ln\left(-\frac{t}{\,m^2\,}\right)\right]
       +2\ln\left(\frac{-u_1}{\,2m^2\,}\right)
         \ln\left(-\frac{t}{\,m^2\,}\right)
                           \nonumber \\[1mm]
& & \hspace*{3cm}
      {}    \left.
        -\ln^2\left(\frac{-t_1}{\,2m^2\,}\right)
        - 2\Li_2\left(1-\frac{t_1}{u_1}\right)
        - \frac{13}{4}\zeta(2)\right\}
\eea
\bea & & \hspace*{-1.5cm}
D(-k_3,-p-k_3,p-k_2,0,0,m,m) =
           iC_{\epsilon}\;\frac{2}{s_1t_1}
           \left\{\,\frac{1}{\epsilon^2}
          + \frac{2}{\epsilon}\,\left[\,
         \ln\left(\frac{-u_1}{\,2m^2\,}\right)
         -\ln\left(\frac{-t_1}{\,2m^2\,}\right) \right.\right.
                   \nonumber \\[1mm]
& & \hspace*{3.6cm}
   {}    \left. -\ln\left(\frac{-s_1}{\,2m^2\,}\right)\right]
       +2\ln^2\left(\frac{-s_1}{\,2m^2\,}\right)
       +2\ln^2\left(\frac{-t_1}{\,2m^2\,}\right)
                           \nonumber \\[1mm]
& & \hspace*{3.6cm}
   {}   -2\ln^2\left(\frac{-u_1}{\,2m^2\,}\right)
        + 4\Li_2\left(\frac{u_1-s_1}{t_1}\right)
        + 4\Li_2\left(\frac{u_1-t_1}{s_1}\right)
                            \nonumber \\[1mm]
& & \hspace*{3.6cm}
      {}    \left.
        -\, 4\Li_2\left(\frac{u_1-t_1}{s_1}
                     \frac{u_1-s_1}{t_1}\right)
        - \frac{3}{2}\zeta(2)\right\} \\[1mm]
& & \hspace*{-1.5cm}
D(-p,-p+k_1,p+k_3,0,m,m,m) = iC_{\epsilon}\;\frac{1}{t_1 s}\,
               \frac{4}{\lambda}\left\{\,
          \Li_2\left(\frac{\lambda-1}{\lambda+\beta}\right)
        + \Li_2\left(\frac{\lambda-1}{\lambda-\beta}\right)
             \right.   \nonumber\\[1mm]
& & \hspace*{4.6cm}
      {}    \left.
        - \Li_2\left(\frac{\lambda+1}{\lambda+\beta}\right)
          - \Li_2\left(\frac{\lambda+1}{\lambda-\beta}\right)
                                   \right\} \\[3mm]
& & \hspace*{1.01cm}\mbox{where}\phantom{kk}
\lambda\equiv\lambda(s_1,t_1)=\sqrt{
\frac{s_1}{s_1+4m^2}\frac{t_1+4m^2}{t_1}} \nonumber
\quad .
\eea
The five-point functions appearing in the calculation of the
Feynman amplitudes shown in Fig.\ref{BOXES}(e,f) can be
rewritten as linear combinations of the four-point integrals
listed above.

In order to evaluate the Coulomb-singular diagrams shown in
Fig.\ref{BOXES}(b) three five-point tensor integrals of the type
\bea\label{EQ_FIVEPT}
\lefteqn{\hspace*{-2.cm}
E_{0,\alpha,\alpha\beta,\alpha\beta\gamma,\alpha\beta\gamma\delta}
(p_c,p_c-k_1,p_c-k_1-k_2,-p_{\,\overline{c}},0,m,m,m,m)  = } \nonumber
\\[2mm]
&&
\hspace*{-1.0cm}
\mu^{4-n}\;\int\frac{\md^nq}{(2\pi)^n}\;
\frac{1,q_{\alpha},q_{\alpha}q_{\beta},q_{\alpha}q_{\beta}q_{\gamma},
        q_{\alpha}q_{\beta}q_{\gamma}q_{\delta}}{[q^2]\,[(q+p_c)^2-m^2]
        \,[(q+p_c-k_1)^2-m^2]}
                    \nonumber \\
& &\hspace{2.5cm}\times
\frac{1}{
\,[(q+p_c-k_1-k_2)^2-m^2]
\,[(q-p_{\,\overline{c}})^2-m^2]}
\eea
have to be calculated. The two additional integrals can be obtained
from (\ref{EQ_FIVEPT}) by exchanging the photon and gluon momenta.
The Coulomb singularity has been isolated by introducing a small
quark velocity $v$ and the momenta $p_c$ and $p_{\,\overline{c}}$
fulfill the relations $\vec{p}_c+\vec{p}_{\,\overline{c}}=\vec{0}$ and
$|\vec{p}_c-\vec{p}_{\,\overline{c}}| = m v$. Since all loop integrals
which have at least one power of the loop momentum $q$ in the numerator
are Coulomb-finite, the tensor reduction of (\ref{EQ_FIVEPT}) can be
performed at $v=0$. The $\pi^2/v$ singularity is thus contained only
in the scalar five-point integrals $E_0$ and in four- and three-point
functions of the type $D(p_c,p_c-k_1,-p_{\,\overline{c}},0,m,m,m)$
and $C(p_c,-p_{\,\overline{c}},0,m,m)$ which emerge from the tensor
reduction. The integration of the three-point function is straightforward
and we obtain in the static limit $v\to 0$
\beq\label{3P1}
C(p_c,-p_{\,\overline{c}},0,m,m) = -iC_{\epsilon}\;\frac{1}{2m^2}\,\left[\,
\frac{1}{\epsilon} + \frac{\pi^2}{v} -2 \right]
\quad .
\eeq
Alternatively, the Coulomb singularity (as well as the infrared pole) can
be isolated by introducing a small gluon mass $\lambda$. In the limit
$\lambda \to 0$ one finds
\bea
C(p_c,-p_{\,\overline{c}},0,m,m) = -iC_{\epsilon}\;\frac{1}{2m^2}\,\left[\,
\ln\left(\frac{\lambda^2}{m^2}\right)+\frac{2\pi}{\lambda}m-2\right]
\eea
which shows the correspondence between the different regularization
schemes (cf.\ Ref.\cite{JK93}):
\bea
\hspace*{3.5cm}\ln\left(\frac{\lambda^2}{m^2}\right) & \Longleftrightarrow &
\frac{1}{\epsilon}\quad\quad\hspace{2mm} \mbox{(IR-singularity)}\nonumber\\
\hspace*{3.5cm}\frac{2\pi}{\lambda}m & \Longleftrightarrow & \frac{\pi^2}{v}
\quad\quad \mbox{(Coulomb-singularity)}\quad .
\eea
The singular piece of the four- and five point functions is determined
by the infrared part of the integration region, $q\to 0$, and can thus
be extracted by neglecting the loop momentum in the infrared finite
propagators. In the limit $|\vec{p}_c-\vec{p}_{\,\overline{c}}| = mv \to 0$
one finds for example
\beq\label{EQ_FPTSING}
D^{(\mbox{\scriptsize IR})}(p_c,p_c-k_1,-p_{\,\overline{c}},0,m,m,m) =
\frac{1}{[(p_c-k_1)^2-m^2]}\;
\frac{(-iC_\epsilon)}{\,2m^2}\left\{\frac{1}{\epsilon}
+\frac{\pi^2}{v} - 2 \right\}
\eeq
and similar expressions for the other four- and five-point functions.
No terms linear in $v$ appear in the relativistic expansion of the
coefficients multiplying the Coulomb-singular integrals. The corresponding
expressions can thus be evaluated in the static limit $v=0$ from the start.

For the finite part of the four-point integral we obtain
\bea\label{FPFIN}
& & \hspace*{-2cm}
D^{(\mbox{\scriptsize fin})}(p_c,p_c-k_1,-p_{\,\overline{c}},0,m,m,m)
\equiv \nonumber\\[1mm]
     & &
D(p_c,p_c-k_1,-p_{\,\overline{c}},0,m,m,m) - \frac{1}{[(p_c-k_1)^2-m^2]}\;
C(p_c,-p_{\,\overline{c}},0,m,m) = \nonumber \\[1.5mm]
     & & \hspace{2cm}
    iC_{\epsilon}\;\frac{2}{t_1m^2}\,\left[\,
      \frac{1}{\beta(t)}\ln\left(\frac{1-\beta(t)}{1+\beta(t)}\right)
    - \ln\left(\frac{2m^2}{t_1}\right) \right]
\quad .
\eea
%The result for $D(p_c,p_c-k_1-k_2,-p_{\,\overline{c}},0,m,m,m)$ can be
%obtained from (\ref{FPFIN}) by interchanging $s_1\leftrightarrow t_1$.
Similarly, the finite part of the five-point integral is given by
\bea & & \hspace*{-2cm}
E^{(\mbox{\scriptsize fin})}(p_c,p_c-k_1,p_c-k_1-k_2,p_{\,\overline{c}},
0,m,m,m,m) \equiv \nonumber\\[1mm]
     & & \hspace*{-1cm}
E(p_c,p_c-k_1,p_c-k_1-k_2,p_{\,\overline{c}},0,m,m,m,m) \nonumber \\
& & \hspace*{1cm}
- \frac{1}{[(p_c-k_1)^2-m^2][(p_c-k_1-k_2)^2-m^2]}\;
C(p_c,-p_{\,\overline{c}},0,m,m)
\quad .
\eea
The difference $E^{(\mbox{\scriptsize fin})}=E-E^{(\mbox{\scriptsize IR})}$
is infrared- and Coulomb-finite and a complete analytical
result has been obtained. Since the corresponding expression is rather
extensive we will only outline the basic steps of the calculation.  By
using the Feynman parametrization technique one obtains
\beq
E = {}-\frac{i}{m^6}\;\left(\frac{\mu^2}{m^2}\right)^{\epsilon}
        \;\frac{\Gamma(3+\epsilon)}{(4\pi)^{n/2}}\;\;{\cal{I}}
\quad ,
\eeq
where
\bea
{\cal{I}} & = &
\int_0^1\,\md\hat{x}\int_0^{1-\hat{x}}\,\md\hat{y}
\int_0^{1-\hat{x}-\hat{y}}\,\md\hat{z}\int_0^{1-\hat{x}-\hat{y}-\hat{z}}
\,\md\hat{w}\;
\left[(\hat{x}-\hat{y}-\hat{z}-\hat{w})^2 -\hat{s}(1-\hat{y}
             -\hat{z})\hat{y}
             \vphantom{x^2}     \right.    \nonumber\\[1mm]
 & & \quad
  \left.{}
-\hat{t}(1+\hat{x}-\hat{y}-\hat{z}-\hat{w})(\hat{y}+\hat{z})
-\hat{u}(1+\hat{x}-\hat{y}-\hat{z}-\hat{w})\hat{y}
\right]^{-3-\epsilon}
\eea
and
\beq
\hat{t} \equiv -\frac{2p\cdot k_1}{m^2}\,, \phantom{kkkk}
\hat{u} \equiv -\frac{2p\cdot k_2}{m^2}\,, \phantom{kkkk}
\hat{s} \equiv  \frac{2k_1\cdot k_2}{m^2}
\quad .
\eeq
Substituting
\beq
\hat{x} = x+y-1\,, \phantom{kkkk}
\hat{y} = w\,, \phantom{kkkk}
\hat{z} = z-w\,, \phantom{kkkk}
\hat{w} = x-z\,,
\eeq
we arrive at an expression which does not depend on the integration
variable $x$:
\beq
{\cal{I}} =
\int_0^1\,\md{x}\int_{1-x}^{2-2{x}}\,\md{y}
\int_0^{{x}}\,\md{z}\int_0^{{z}}\,\md{w}\;
\left[(1-y)^2 - \hat{t}yz -\hat{u}yw -\hat{s}w(1-z) \right]^{-3-\epsilon}
\quad .
\eeq
Interchanging the order of integration and evaluating two integrations,
we find
\beq
{\cal{I}} = \frac{1}{2+\epsilon}\int_0^1\,\md{z}\int_0^z\,\md{y}\,
(y-z)\left[f(\epsilon,\hat{s},\hat{t},\hat{u},z,y)
    - 2f(\epsilon,\hat{s},\hat{t},\hat{u},z,2y)\right]
\quad ,
\eeq
where
\bea
f(\epsilon,\hat{s},\hat{t},\hat{u},z,y) & \equiv & \frac{1}{\hat{u}y
+\hat{s}z}\left\{
\left[(1-y)^2-(1-z)(y(\hat{t}+\hat{u})+\hat{s}z)\right]^{-2-\epsilon}
\right.
\nonumber\\[1mm]
&&\left. \hspace{2cm} -
\left[(1-y)^2-\hat{t}y(1-z)\right]^{-2-\epsilon} \right\}
\quad .
\eea
The singularity is contained in the expression
\beq
{\cal{I}}^S = \frac{1}{2+\epsilon}\int_0^1\,\md{z}\int_0^z\,\md{y}\,
(y-1)\,f^S(\epsilon,\hat{s},\hat{t},\hat{u},z,y)
\eeq
with
\bea
f^S(\epsilon,\hat{s},\hat{t},\hat{u},z,y) & \equiv & \frac{1}{\hat{u}
+\hat{s}}\left\{
\left[(1-y)^2-(1-z)(\hat{t}+\hat{u}+\hat{s})\right]^{-2-\epsilon}
\right.\nonumber\\[1mm]
&&\left.\hspace{3cm} -
\left[(1-y)^2-\hat{t}(1-z)\right]^{-2-\epsilon} \right\}
\quad .
\eea
The integration of ${\cal{I}}^S$ is straightforward and we obtain
\bea
{\cal{I}}^S & = &
-\frac{1}{(1+\epsilon)}\frac{1}{(2+\epsilon)}
\left[-\frac{1}{\hat{t}(\hat{s}+\hat{t}+\hat{u})}\frac{1}{2\epsilon}
+\frac{1}{\hat{s}+\hat{u}}\left\{\frac{1}{\hat{t}}
\ln\left(\frac{-\hat{t}}{1-\hat{t}}\right)
\right.\right.\nonumber\\[1mm]
&& \left.\left.\hspace{2cm}
-\frac{1}{\hat{s}+\hat{t}+\hat{u}}
\ln\left(\frac{-\hat{s}-\hat{t}-\hat{u}}{1-\hat{s}-\hat{t}-\hat{u}}
\right)\right\}\right]
\quad .
\eea
The remaining difference
\bea
{\cal{I}}^{\,\mbox{\scriptsize{}fin}} & = &
\frac{1}{2}\int_0^1\,\md{z}\int_0^z\,\md{y}
\left[(y-z)\left\{f(\epsilon,\hat{s},\hat{t},\hat{u},z,y)
    + 2f(\epsilon,\hat{s},\hat{t},\hat{u},z,2y)\right\}
\right.\nonumber\\[1mm]
&&\left.\hspace{3cm}
-(y-1)\,f^S(\epsilon,\hat{s},\hat{t},\hat{u},z,y) \right]
\eea
is finite and can be evaluated in the limit $\epsilon\to 0$ by
using elementary integration techniques.

\section{The three-particle final states}\label{APP2}

In this Appendix we outline the kinematics for the processes
involving  three-particle final states and give explicit results
for the infrared-collinear cross section.

\subsection{The phase space integration}\label{APP2A}
For the calculation of the two-to-three-body processes
(\ref{EQ_TWOGL}), (\ref{EQ_TWOQ}) and (\ref{EQ_GAMQ})
we closely follow Ref.\cite{BKVS89} and
introduce the following 10 kinematical invariants:\\
\parbox{\textwidth}{
\parbox{7cm}{
\begin{eqnarray*}
s   &\!\! = \!\!& (k_1+k_2)^2 \equiv s_1 + 4m^2     \\
s_3 &\!\! = \!\!& (k_3+k_4)^2            \\
s_4 &\!\! = \!\!& (2p+k_4)^2 - 4m^2      \\
s_5 &\!\! = \!\!& (2p+k_3)^2 - 4m^2      \\
t_1 &\!\! = \!\!& (2p-k_1)^2 - 4m^2
\end{eqnarray*}}
\parbox{7cm}{
\begin{eqnarray*}
t_6 &\!\! = \!\!& (k_3-k_2)^2        \\
u_6 &\!\! = \!\!& (k_3-k_1)^2        \\
u_1 &\!\! = \!\!& (2p-k_2)^2 - 4m^2      \\
t'  &\!\! = \!\!& (k_4-k_1)^2        \\
u'  &\!\! = \!\!& (k_4-k_2)^2
\quad .
\end{eqnarray*}}
\hfill\parbox{1cm}{\bea\label{JPREALINV}\eea}}
where $k_1+k_2 = 2p + k_3 + k_4$. The invariants $s$, $t_1$ and $u_1$
have already been defined in (\ref{INVDEF1}). Since we are dealing with a
three-particle final state only five of the invariants are linearly
independent. In the centre of mass frame of the outgoing light particles
(gluons and/or light (anti-)quarks), the momenta are given by
\bea
k_1 & = & (\omega_1,\dots,0,0,\omega_1) \nonumber \\
P_{J/\psi} = 2p & = & (E_{J/\psi},\ldots,0,p_{J/\psi}\sin\psi,
                       p_{J/\psi}\cos\psi) \nonumber \\
k_3 & = & (\omega_3,\ldots,-\omega_3\sin\theta_1\sin\theta_2,
               -\omega_3\sin\theta_1\cos\theta_2,
               -\omega_3\cos\theta_1) \nonumber \\
k_4 & = & (\omega_4,\ldots,\omega_3\sin\theta_1\sin\theta_2,
               \omega_3\sin\theta_1\cos\theta_2,
               \omega_3\cos\theta_1) \nonumber \\
k_2 & = & (\omega_2,\ldots,0, p_{J/\psi}\sin\psi,p_{J/\psi}\cos\psi
          -\omega_1)
\eea
The coordinate axes can be chosen in such a way that the $n-4$
unspecified angular components do not contribute to the matrix
element squared. From four-momentum conservation and the on-mass-shell
constraints one can derive the identities
\bea
%\lefteqn{\hspace*{-3.6cm}
&&
\omega_1 = \frac{s_1+t_1+4m^2}{2\,\sqrt{s_3\vphantom{l}}}, \phantom{kkkk}
\omega_2 = \frac{s_1+u_1+4m^2}{2\,\sqrt{s_3\vphantom{l}}}, \phantom{kkkk}
\omega_3 = \omega_4 = \frac{\sqrt{s_3\,\vphantom{l}}}{2}
 \nonumber \\
&&E_{J/\psi} = \frac{s_1-s_3}{2\,\sqrt{s_3\vphantom{l}}}, \phantom{kkkk}
p_{J/\psi} = \sqrt{E_{J/\psi}^2-4m^2}, \phantom{kkkk}
%\\ \vphantom{\overbrace{\frac{\omega_1^2-\omega_2^2
%          +p_{J/\psi}^2}{2p_{J/\psi}\omega_1}}}
\cos\psi  =
\frac{\omega_1^2-\omega_2^2+p_{J/\psi}^2}{2\,p_{J/\psi}\omega_1}
\quad .
\eea
For the cross section one has
\bea\label{SIGTOTR}
\sigma & = & \frac{1}{2!}\;K_C\,K_S\;\frac{1}{2s}\;
\mu^{2(4-n)}\;
\int\frac{\md^nP_{J/\psi}}{(2\pi)^{n-1}}\,
\int\frac{\md^nk_3}{(2\pi)^{n-1}}\,
\int\frac{\md^nk_4}{(2\pi)^{n-1}}\,
\delta^+(P_{J/\psi}^2-4m^2)
 \nonumber \\[1mm]
& & \times
\delta^+(k_3^2)\,
\delta^+(k_4^2)\;\,
(2\pi)^n\,\delta^n(k_1+k_2-P_{J/\psi}-k_3-k_4)
\;\,|{\cal{M}}^R|^2 \; ,
\eea
$K_C$ and $K_S$ denoting the colour and spin averaging factors. The
factor $1/2!$ has to be included for the two-gluon final states
because of Bose symmetry. It is convenient to rewrite the cross
section according to
\bea\label{PSX}
\lefteqn{\hspace*{-0.9cm}
\sigma = K_C\,K_S\;\frac{1}{2s}\;
\frac{\mu^{2(4-n)}}{(2\pi)^{2n-3}}\;
\int \md^nP_{J/\psi}\,d^n\xi\;
\delta^+(P_{J/\psi}^2-4m^2)\,
\delta^n(k_1+k_2-P_{J/\psi}-\xi)
} \nonumber \\
\hspace*{-1.3cm}
& & \times
\frac{1}{2!}\;
\int \md^nk_3\,
\int \md^nk_4\,
\delta^+(k_3^2)\,
\delta^+(k_4^2)\;
\delta^n(\xi-k_3-k_4)
\;\,|{\cal{M}}^R|^2 \phantom{kkk}
\label{SIGTOTR2}
\quad .
\eea
The integral over the phase space of the two outgoing light partons
is evaluated in the corresponding centre of mass frame:
\bea
\lefteqn{\hspace*{-0.5cm}
\frac{1}{2!}\;
\int \md^nk_3\,
\int \md^nk_4\,
\delta^+(k_3^2)\,
\delta^+(k_4^2)\;
\delta^n(\xi-k_3-k_4)
\;\,\mid\!M_R\!\mid^2
\;= }  \\[1mm]
\hspace*{-2.3cm}
& &
s_3^{(n-4)/2}
\;\pi^{(n-4)/2}
\;\frac{1}{4}\,
\frac{\Gamma(\frac{n}{2}-1)}{\Gamma(n-3)}
\,\frac{1}{2}\,
\int_0^{\pi}\md\theta_1\sin^{n-3}\theta_1
\int_0^{\pi}\md\theta_2\sin^{n-4}\theta_2
\;\,|{\cal{M}}^R|^2 \nonumber
\quad .
\eea
The remaining integrations are performed in the centre of mass frame
of the initial state particles with momenta given by
\bea
k_1 & = & \sqrt{s}/2\,(1,\ldots,0,0,1) \nonumber \\
k_2 & = & \sqrt{s}/2\,(1,\ldots,0,0,-1) \nonumber  \\
P_{J/\psi} & = & (E_{J/\psi},\ldots,0,p_{J/\psi}\sin\chi,p_{J/\psi}
                 \cos\chi) \quad .
\eea
One finds for the cross section
\bea
\sigma & = & K_C\,K_S\;\frac{1}{s}\;
\frac{(4\pi)^{-n}\,\mu^{2(4-n)}}{\Gamma(n-3)}\;
\int \md{}E_{J/\psi}\,(E_{J/\psi}^2-4m^2)^{\frac{n-3}{2}}\,
\int_0^{\pi}\md\chi\sin^{n-3}\chi
\nonumber \\
& & \phantom{kkkk}\times
s_3^{(n-4)/2}\;\frac{1}{2}\;
\int_0^{\pi}\md\theta_1\sin^{n-3}\theta_1
\int_0^{\pi}\md\theta_2\sin^{n-4}\theta_2
\;\,|{\cal{M}}^R|^2
\label{SIGTOTR3}
\quad .
\eea
After changing the integration variables
$(E_{J/\psi},\chi)\to (t_1,u_1)$ using
\beq
E_{J/\psi} = - \frac{t_1+u_1}{2\sqrt{s}}, \phantom{kkkkkk}
\cos\chi = \frac{t_1-u_1}{\sqrt{(t_1+u_1)^2-16\,m^2s}}
\eeq
we finally obtain
\bea
s^2\frac{\md^2\sigma}{\md t_1\md u_1} & = & K_C\,K_S\,
\frac{(4\pi)^{-n}\,\mu^{4-n}}{2\Gamma(n-3)}\;
\left(
\frac{t_1u_1-4m^2s}{\mu^2s}\right)^{(n-4)/2}
\nonumber \\
& &
\times
s_3^{(n-4)/2}\;\frac{1}{2}\;
\int_0^{\pi}\md\theta_1\sin^{n-3}\theta_1
\int_0^{\pi}\md\theta_2\sin^{n-4}\theta_2
\;\,|{\cal{M}}^R|^2
\label{SIGTOTRFIN}
\;\; .
\eea
The kinematical limits on $t_1, u_1$ can be deduced from
the constraints $-1\le\cos\chi\le 1$ and $s_3\ge 0$:
\beq
-s \le t_1 \le -4m^2, \phantom{kkkkk}
-s_1-t_1-8m^2 \le u_1 \le \frac{4m^2s}{t_1}
\quad .
\eeq

To perform the angular integrations the matrix element squared has
been decomposed into sums of terms which have at most two factors
containing the dependence on the polar angle $\theta_1$ and the
azimuthal angle $\theta_2$. This partial fractioning exploits the
kinematical relations between the invariants defined in
(\ref{JPREALINV}):
\bea
s_3 &\!\! = \!\!& - u_6 - t_6 - s_5  \nonumber \\
s_4 &\!\! = \!\!&   s_1 + u_6 + t_6    \nonumber \\
s_5 &\!\! = \!\!& - t_1 - u_1 - s_4 - 8m^2 \nonumber \\
t_6 &\!\! = \!\!& - u_1 - s_1 - u' - 4m^2 \nonumber \\
u_6 &\!\! = \!\!& - t_1 - s_1 - t' - 4m^2 \quad .
\eea
By using identities as e.g.
\beq
\frac{1}{s_5t_6(s_4+t_1+2t')} = \frac{1}{u_1}\left(\frac{1}{s_5t_6}
-\frac{1}{t_6(s_4+t_1+2t')}-\frac{2}{s_5(s_4+t_1+2t')}\right)
\eeq
the angular integration can be reduced to the evaluation of standard
integrals of the form
\bea\label{EQ_ANGINT}
I_{n}^{(k,l)} & = &
 \int_0^\pi\! \md\theta_1\sin^{n-3}\theta_1
 \int_0^\pi\! \md\theta_2\sin^{n-4}\theta_2
\nonumber\\
&& \hspace{2cm}
\times (a+b\cos\theta_1)^{-k}
(A+B\cos\theta_1+C\sin\theta_1\cos\theta_2)^{-l}
\quad .
\eea
The results for these integrals can be found in Ref.\cite{BKVS89}.

\subsection{The infrared-collinear cross section}\label{APP2B}
The calculation of the infrared-collinear cross section
(\ref{EQ_JPSOFT}) will be outlined in this Appendix. The resulting
expressions contain all divergences due to soft gluon emission and
splitting of the final state gluon into gluon and light
quark-antiquark pairs.

We begin by considering the infrared region of phase space
where one of the final state gluon momenta $k_{3,4}$ becomes
soft. In the limit $k_4\to 0$ the amplitude of the real gluon
emission factorizes into an eikonal factor multiplying the
Born amplitude:
\beq\label{EQ_JPMSQSOFT1}
\left.\mid\!{\cal{M}}^R\!\mid^2\right|_{k_4\to 0} =
\frac{-2k_2\cdot k_3}{s_3u'}\;g^2\frac12
(f^{abc})^2\left.\mid\!{\cal{M}}^B\!\mid^2\right|_{Tr^2(T^aT^b) = 1}
\quad ,
\eeq
and analogous for $k_3\to 0$
\beq\label{EQ_JPMSQSOFT2}
\left.\mid\!{\cal{M}}^R\!\mid^2\right|_{k_3\to 0} =
\frac{-2k_2\cdot k_4}{s_3t_6}\;g^2\frac12
(f^{abc})^2\left.\mid\!{\cal{M}}^B\!\mid^2\right|_{Tr^2(T^aT^b) = 1}
\quad ,
\eeq
where $(f^{abc})^2 = N(N^2-1)$.

In addition to the soft gluon divergences, the amplitudes for the
$2\to 3$ processes contain terms which are singular when the two
outgoing light partons become collinear, $k_3\| k_4$, i.e.
$\cos\theta_{34} \equiv \cos\theta = 1$. For the calculation of the
collinear matrix element we choose the four-momenta in the
centre of mass frame of the incoming partons according to
\bea\label{EQ_COLLMOM}
k_1 & = & \frac{\sqrt{s}}{2}\,(1,\ldots,\sin\chi,0,\cos\chi)
\nonumber\\
k_2 & = & \frac{\sqrt{s}}{2}\,(1,\ldots,-\sin\chi,0,-\cos\chi)
\nonumber\\
k_3 & = & \omega_3\,(1,\ldots,0,0,1)
\nonumber\\
k_4 & = & \omega_4\,(1,\ldots,\sin\theta\cos\phi,
                     \sin\theta\sin\phi,\cos\theta)
\nonumber\\
P = 2p & = & (E,\ldots,-\omega_4\sin\theta\cos\phi,
             -\omega_4\sin\theta\sin\phi,-\omega_4\cos\theta-\omega_3)
\quad .
\eea
Hence, in the limit $\theta\to 0$ the condition $s_3<\Delta$ is
given by
\beq\label{EQ_IRCPS}
s_3 = 2\,\omega_3\omega_4\,(1-\cos\theta)
\approx \omega_3\omega_4\,\theta^2 < \Delta \quad .
\eeq
When integrated over the solid angle
\beq\label{EQ_SA}
\md\Omega \propto \md\theta\sin^{n-3}\theta
\approx \theta^{n-3}\md\theta
\quad ,
\eeq
the $1/s_3 \sim 1/\theta^2$ terms in the matrix element squared lead
to logarithmic singularities in the limit $\theta\to 0$. The pieces
with additional powers of $\theta$ in the numerator only give
contributions of the ${\cal{O}}(\Delta)$ which vanish in the limit
$\Delta\to 0$.  Accordingly, all invariants in the expressions
multiplying the $1/s_3$ propagator can be replaced by their values at
$\theta = 0$:\\
\parbox{\textwidth}{
\parbox{7cm}{
\begin{eqnarray*}
s_3^0 &\!\! = \!\!& 0        \\
s_4^0 &\!\! = \!\!& z s_1    \\
s_5^0 &\!\! = \!\!& (1-z)s_1 \\
t'^0 &\!\! = \!\!& z(u_1+4m^2)
\end{eqnarray*}}
\parbox{7cm}{
\begin{eqnarray*}
u'^0 &\!\! = \!\!& z(t_1+4m^2)     \\
t_6^0 &\!\! = \!\!& (1-z)(t_1+4m^2) \\
u_6^0 &\!\! = \!\!& (1-z)(u_1+4m^2)
\quad .
\end{eqnarray*}}
\hfill\parbox{1cm}{\bea\label{EQ_INVT0}\eea}}
The other invariants remain unaltered. The splitting parameter $z$ is
defined as
\beq
z = \frac{\omega_4}{\omega_3+\omega_4}
\quad
\eeq
and should not be confused with the $J/\psi$ energy variable defined
in (\ref{ZDEF}). The parts of the matrix element squared which are
proportional to $1/s_3^2\sim1/\theta^4$ have to be expanded up to the
${\cal{O}}(\theta^2)$.  Gauge invariance ensures that in the resulting
expression all terms $\sim 1/\theta^4$ cancel. The contributions $\sim
\theta/\theta^4$ are proportional to $\cos\phi$ and vanish after the
azimuthal integration. The parts $\sim \theta^2/\theta^4$ finally
contribute to the collinear matrix element and lead to logarithmic
singularities when integrated over the solid angle (\ref{EQ_SA}).

For the process $\gamma g \to J/\psi + g g $ we obtain the following
result for the matrix element squared in the limit $\theta\to 0$:
\bea\label{EQ_MSQCOLL}
\lefteqn{
\int_0^\pi\md\phi\sin^{n-4}\phi\left.|{\cal{M}}^R|^2\right|_{\theta\to 0}
= }\nonumber\\[1mm]
&&B\left({\textstyle{}\frac{n-3}{2};\frac12}\right)
\frac{1}{6}P_{gg}(z) \frac{1}{s_3}\;
g^2(f^{abc})^2\left.\mid\!{\cal{M}}^B\!\mid^2\right|_{Tr^2(T^aT^b) = 1}
+{\cal{O}}\left(\frac{1}{\theta}\right)
\quad ,
\eea
where $B(x;y)=\Gamma(x)\Gamma(y)/\Gamma(x+y)$ is the Beta-function.
The gluon-gluon splitting function $P_{gg}$ is given by
\beq
P_{gg} = 6\left[\frac{1-z}{z}+\frac{z}{1-z}+z(1-z)\right]\quad .
\eeq
As mentioned before, the terms of the order $1/\theta$ which have
not been included in (\ref{EQ_MSQCOLL}) only give contributions of the
${\cal{O}}(\Delta)$ and thus vanish in the limit $\Delta\to 0$.

\vspace{2mm}

\begin{figure}[hbtp]

\vspace*{0.25cm}
\parbox[t]{6.0cm}{
\hspace*{0.5cm}
\epsfig{%
file=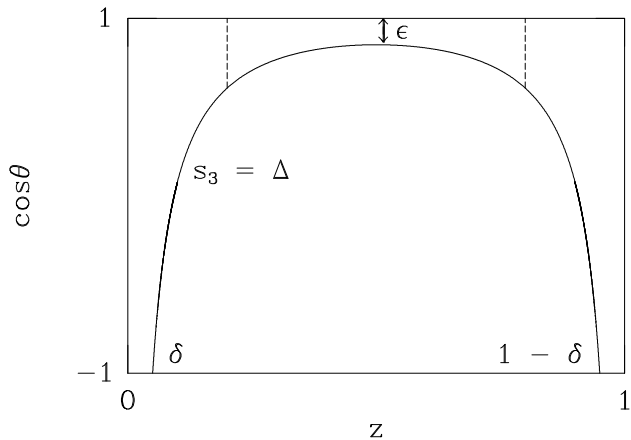,%
height=9cm,%
width=12.4cm,%
bbllx=1.0cm,%
bblly=1.9cm,%
bburx=19.4cm,%
bbury=26.7cm,%
rheight=8.2cm,%
rwidth=15cm,%
angle=-90}
}
\vspace*{-2.45cm}
\hfill\parbox[t]{5.1cm}{
\vspace*{-8.45cm}
\caption[ggdiag]{\label{SCREGION}
                 The in\-fra\-red-col\-li\-ne\-ar region of phase
                 space in the $z$-$\theta$-plane. The infrared
                 parameter $\delta$ and the collinear parameter
                 $\epsilon$ are obtained from  $s_3=\Delta$
                 with $s_3\sim z\,(1-z)\,(1-\cos\theta)$ and
                 vanish when $\Delta\to 0$. } }

\end{figure}

The infrared-collinear region of phase space (\ref{EQ_IRCPS})
is illustrated in the $z$-$\theta$-plane in Fig.\ref{SCREGION}.
%-----------------------------------------------------------------------
%-----------------------------------------------------------------------
For $k_4\to 0$ ($z\to 0$) and $k_3\to 0$ ($z\to 1$), the matrix
element squared can be approximated by the expressions
(\ref{EQ_JPMSQSOFT1}) and (\ref{EQ_JPMSQSOFT2}), respectively,
while for $\theta\to 0$ the collinear limit (\ref{EQ_MSQCOLL})
has to be used. It is clear from (\ref{EQ_JPMSQSOFT1}),
(\ref{EQ_JPMSQSOFT2}) and (\ref{EQ_MSQCOLL}) that in the
infrared-collinear region, i.e. in the upper left and right
corner in Fig.\ref{SCREGION}, both approximations are identical.

In order to obtain a result for the matrix element squared in the
region $s_3<\Delta$ which is Lorentz-frame independent, the splitting
function $P_{gg}$ of the collinear cross section can be rewritten in
terms of invariant variables according to
\bea\label{EQ_PGGINV}
\frac{1}{6}P_{gg}(z)& = &\left[\frac{1-z}{z}+\frac{z}{1-z}+z(1-z)
\right] \nonumber\\[1mm]
&&\hspace{2cm} =\frac{k_2\cdot k_3}{k_2\cdot k_4}
              + \frac{k_2\cdot k_4}{k_2\cdot k_3} +
\frac{(p\cdot k_3)(p\cdot k_4)}{[p\cdot (k_3+k_4)]^2} + {\cal{O}}(\theta)
\nonumber\\
&&\hspace{2cm} =\frac{t_6}{u'}+\frac{u'}{t_6}+\frac{s_4s_5}{(s_4+s_5)^2}
+ {\cal{O}}(\theta)\quad .
\eea
The expression (\ref{EQ_PGGINV}) is Lorentz-invariant and can be used
as an eikonal factor not only for the collinear configuration but also
for the infrared region.  Note that no additional divergences have
been introduced in (\ref{EQ_PGGINV}) since for $k_2 \| k_4$ and $k_2
\| k_3$ also the corresponding numerators vanish in the limit
$\theta\to 0$, $k_2 \| k_3$ and $k_2 \| k_4$, respectively.
Alternative ways of rewriting the splitting function in terms of
invariants as e.g.
\beq
\frac{1-z}{z} = \frac{p\cdot k_3}{p\cdot k_4} + {\cal{O}}(\theta)
\eeq
only approximate the collinear matrix element squared but can
not be used to describe the infrared region.

An explicit distinction between collinear and infrared configurations
as has been introduced in the centre of mass frame of the incoming
partons according to (\ref{EQ_COLLMOM}) is no longer necessary. Using
the invariant formulation for the matrix element squared in the region
$s_3<\Delta$ as given by the eikonal factor (\ref{EQ_PGGINV}), the
phase space integrations necessary for the calculation of the
corresponding cross section can conveniently be performed the
centre of mass frame of the outgoing light partons:
\bea\label{EQ_S3MSQ}
\lefteqn{\int_{0}^{\Delta}\md s_3\, s_3^{-\epsilon}\int
\md\Omega_n\sum \left|{\cal{M}}^R\right|^2} \nonumber \\[2mm]
&&=\int_{0}^{\Delta}\md s_3 \,s_3^{-\epsilon}\int
\md\theta\sin^{n-3}\theta
\; g^2(f^{abc})^2 \nonumber \\[1mm]
&&\hphantom{=\int_{0}^{\Delta}\md s_3 s_3^{-\epsilon}}\times
{\textstyle{}B\left(\frac{n-3}{2};\frac12\right)}
\frac{1}{s_3}\left(
%\frac{k_2k_3}{k_2k_4} + \frac{k_2k_4}{k_2k_3} +
%\frac{(k_2k_3)(k_2k_4)}{[k_2(k_3+k_4)]^2}
\frac{t_6}{u'}+\frac{u'}{t_6}+\frac{s_4s_5}{(s_4+s_5)^2}
\right)\sum\left.\mid\!{\cal{M}}^B\!\mid^2\right|_{Tr^2(T^aT^b) = 1}
\nonumber\\[2mm]
&&=2\pi\, g^2\,(f^{abc})^2 \,
\frac{\Gamma(n-3)}{[\Gamma((n-2)/2)]^2}
\left\{2B\left(\frac{n}{2};\frac{4-n}{2}\right)
+B\left(\frac{4-n}{2},\frac{n}{2}\right)\right\}\nonumber\\[1mm]
&&\hphantom{=\int_{0}^{\Delta}\md s_3 s_3^{-\epsilon}}\times
\sum\left.\mid\!{\cal{M}}^B\!\mid^2\right|_{Tr^2(T^aT^b) = 1}\,\times\,
\int_{0}^{\Delta}\md s_3 s_3^{-1-\epsilon}
\nonumber\\[2mm]
&&=g^2(f^{abc})^2 \;m^{2\epsilon}\,4\pi\left(\frac{\Delta}{m^2}
\right)^{-\epsilon}\left[\frac{1}{\epsilon^2}+\frac{1}{\epsilon}
\left(1-\frac{1}{12}\right)+2-\frac{5}{36}\right]
\sum\left.\mid\!{\cal{M}}^B\!\mid^2\right|_{Tr^2(T^aT^b) = 1}
\quad .
\eea
By using (\ref{EQ_JPSOFT}) and (\ref{EQ_S3MSQ}) one finally obtains
for the cross section of the process $\gamma g \to J/\psi + g g$ in
the region $s_3 < \Delta$:
\bea\label{EQ_MSQCOLL1}
s^2 \left.\frac{\md\sigma}{\md t_1\md u_1}\right|_{s_3\le\Delta}
&\!\! =\!\! & \frac{1}{(N^2-1)}
\frac{1}{4(1-\epsilon)^2}
\frac{\pi (4\pi)^{-2+\epsilon}}{\Gamma(1-\epsilon)}
\left(\frac{t_1u_1-4m^2s}{\mu^2s}\right)^{-\epsilon}
\!\delta(s_1+t_1+u_1+8m^2)\nonumber\\[1mm]
&& \times C_{\epsilon}\,g^2\, (f^{abc})^2\,
\left[\frac{1}{\epsilon^2}+\frac{1}{\epsilon}\left(1-\frac{1}{12}
-\ln\frac{\Delta}{m^2}\right)+2 -\frac{5}{36}\right.
\nonumber\\[1mm]
&&\left.
+\frac12\ln^2\frac{\Delta}{m^2}-\frac32\zeta(2)-
\left(1-\frac{1}{12}\right)\ln\frac{\Delta}{m^2}\right]\left.
|{\cal{M}}^{B}|^{2}\right|_{Tr^2(T^aT^b) = 1}
\eea
with $C_{\epsilon}$ as defined in (\ref{EQ_CEPS}).

The collinear matrix element squared for the process $\gamma g \to J/\psi
+ q \overline{q}$ is found to be
\bea\label{EQ_MSQCOLL2}
\lefteqn{
\int_0^\pi\md\phi\sin^{n-4}\phi\left.|{\cal{M}}^R|^2\right|_{\theta\to
0} = } \nonumber\\
&&={\textstyle{}B\left(\frac{n-3}{2};\frac12\right) }
P_{qg}(z) \frac{2}{s_3}\;
g^2\left.\mid\!{\cal{M}}^B\!\mid^2\right|_{Tr^2(T^aT^b) = 1}
+{\cal{O}}\left(\frac{1}{\theta}\right)
\quad ,
\eea
with the gluon-(anti)quark splitting function $P_{qg}$
\beq
P_{qg} = P_{\overline{q}g} = \frac{z^2 + (1-z)^2-\epsilon}{1-\epsilon}
\quad .
\eeq
By using the invariant expression
\beq
P_{qg} = P_{\overline{q}g} = \left[\left(
\frac{s_4}{s_1}\right)^2 +
\left(\frac{s_5}{s_1}\right)^2 -\epsilon\right]\frac{1}{1-\epsilon}
+ {\cal{O}}(\theta)
\eeq
we find for the cross section of the process
$\gamma g \to J/\psi + q\overline{q}$
in the region $s_3 < \Delta$:
\bea
s^2 \left.\frac{\md\sigma}{\md t_1\md u_1}\right|_{s_3\le\Delta}
&\!\! =\!\! & \frac{1}{(N^2-1)}\frac{1}{4(1-\epsilon)^2}
\frac{\pi (4\pi)^{-2+\epsilon}}{\Gamma(1-\epsilon)}
\left(\frac{t_1u_1-4m^2s}{\mu^2s}\right)^{-\epsilon}
\!\!\delta(s_1+t_1+u_1+8m^2) \nonumber\\[1mm]
&&\times \nlf\;C_{\epsilon}\;g^2\,\frac43
\left(-\frac{1}{\epsilon}\right)\left[1+\frac53\epsilon
- \epsilon\ln\frac{\Delta}{m^2}\right]\left.
|{\cal{M}}^{B}|^2\right|_{Tr^2(T^aT^b) = 1}\; .
\eea
It has been checked explicitly that an expansion of the full matrix
element squared in terms of small $s_3$ leads to results identical
to the equation above.

\section{The kinematics of the photon-hadron cross section}\label{APP3}
In this Appendix we discuss the kinematics of the hadronic differential
and total cross section. The single-particle inclusive hadronic cross
section for the process
\beq
\gamma(K_1) + P(P_1) \to J/\psi(2p) + X
\eeq
reads
\beq
S^2 \frac{\md\sigma^{\gamma P}}{\md T_1\md U_1} =
\sum_{i=g,q(\overline{q})}\; \int_{x_{\min}}^{1}
\frac{\md x}{x} f^{P}_{i}(x,Q^2) \, s^2
\frac{\md\sigma^{\gamma i}}{\md t_1\md u_1}
\quad ,
\eeq
$f^{P}_{g}$ and $f^{P}_{q(\overline{q})}$ denoting the parton
distributions in the proton. The hadronic invariants are defined
according to
\bea\label{HADINV}
S   & \!\! = \!\! & (K_1+P_1)^2 = s/x     \nonumber\\
T_1 & \!\! = \!\! & (K_1-2p)^2-4m^2 = t_1 \nonumber\\
U_1 & \!\! = \!\! & (P_1-2p)^2-4m^2 = u_1/x
\quad .
\eea
{}From (\ref{HADINV}) and the kinematical condition
$xS+T_1+xU_1+4m^2\ge\Delta$ one deduces
\beq
x_{\min} = \frac{\Delta-T_1-4m^2}{S+U_1}
\quad .
\eeq

It is convenient to rewrite the integration in terms of the
$J/\psi$ energy variable $z$ (\ref{ZDEF}) and the $J/\psi$
transverse momentum $p_\perp$. By using the relations
\beq
s = x S \quad\quad
t_1 = -\frac{1}{z}\left(p_{\perp}^{2}+4m^2\right) \quad\quad
u_1 = -xz S
\eeq
one finds
\beq\label{EQ_HADRONZPT}
\frac{\md\sigma^{\gamma P}}{\md p_{\perp}^{2}\md z} =
\sum_{i=g,q(\overline{q})} \; \int_{x_{\min}}^{1} \md x
f^{P}_{i}(x,Q^2)\, \frac{xS}{z} \,\frac{\md\sigma^{\gamma i}}{\md
t_1\md u_1} \quad ,
\eeq
where
\beq
x_{\min} = \frac{z\Delta+p_{\perp}^{2}+4m^2(1-z)}{Sz(1-z)}
\quad .
\eeq
The total hadronic cross section is obtained by integration over
$p_{\perp}^{2}$ and $z$:
\beq\label{EQ_JPPHASE2}
\sigma^{\gamma P}(S,m^2) =
\int_{z_{\min}}^{z_{\max}}\! \md z
\int_{p_{\perp\,\min}^{2}}^{p_{\perp\,\max}^{2}}\!\md
 p_{\perp}^{2} \;\;\frac{\md\sigma^{\gamma P}}{\md
 p_{\perp}^{2}\md z} (S,p_{\perp}^{2},z,m^2) \quad ,
\eeq
with
\bea\hspace*{2cm}
z_{\max} \!\! & = & \!\! 1 \nonumber\\
z_{\min} \!\! & = & \!\! 4m^2/S                             \nonumber\\
p_{\perp\,\max}^{2} \!\! & = & \!\!  (1-z)(Sz-4m^2)-z\Delta \nonumber\\
p_{\perp\,\min}^{2} \!\! & = & \!\!  0
\quad .
\eea
Finally, the $p_{\perp}^{2}$ distribution is given by
\beq
\frac{\md\sigma^{\gamma P}(S,p_{\perp}^{2},m^2)}{\md p_{\perp}^{2}} =
\int_{z_{\min}}^{z_{\max}}\! \md z
\;\;\frac{\md\sigma^{\gamma P}}{\md p_{\perp}^{2}\md z}
(S,p_{\perp}^{2},z,m^2)
\quad ,
\eeq
where
\beq\label{EQ_JPPHASE3}
z_{{\max \atop \min}}  =  \frac{(S+4m^2-\Delta)}{2S}
\left(1\pm\sqrt{1-\frac{4S(4m^2+p_{\perp}^{2})}{(S+4m^2
 -\Delta)^2}}\right) \quad .
\eeq

The cut-off parameter $\Delta$ in the formulas above has to be
included only for the calculation of the hard-gluon cross section. In
our numerical program we have rewritten the $\ln^{i}\Delta (i=1,2)$
singularities of the virtual plus soft cross section as integrals over
the momentum fractions of the partons $x$. The corresponding
contributions have then been added to the hard cross section,
cancelling the equivalent logarithms in this part so that the limit
$\Delta \to 0$ could safely be carried out.

\end{document}